\documentstyle[11pt,twoside,psfig,jltp]{article}

\newcommand{\sty}{|\!\! \uparrow \rangle}
\newcommand{\sta}{|\!\! \downarrow \rangle}
\newcommand{\stya}{|\!\! \uparrow \downarrow \rangle}
\newcommand{\st}{|0\rangle}

\newcommand{\Aa}[1]{A_{#1\downarrow}}
\newcommand{\Ay}[1]{A_{#1\uparrow}}
\newcommand{\Aad}[1]{A_{#1\downarrow}^{\dagger}}
\newcommand{\Ayd}[1]{A_{#1\uparrow}^{\dagger}}
\newcommand{\As}[1]{A_{#1\sigma}}
\newcommand{\Asd}[1]{A_{#1\sigma}^{\dagger}}
\newcommand{\uni}{\hat{1}}

\newcommand{\ds}{d_{\sigma}}
\newcommand{\dms}{d_{-\sigma}}
\newcommand{\dsd}{d_{\sigma}^{\dagger}}
\newcommand{\dmsd}{d_{-\sigma}^{\dagger}}

\newcommand{\nyna}{n_{\uparrow}n_{\downarrow}}
\newcommand{\nsig}{\langle n_{\sigma} \rangle}
\newcommand{\nms}{\langle n_{-\sigma} \rangle}
\newcommand{\nsignms}{\langle n_{\sigma}n_{-\sigma} \rangle}
\newcommand{\nsigb}{n_{\sigma}}
\newcommand{\nmsb}{n_{-\sigma}}
\newcommand{\nup}{\langle n_{\uparrow} \rangle}
\newcommand{\ndown}{\langle n_{\downarrow} \rangle}

\newcommand{\spinsusc}{\chi}
\newcommand{\spinsuscperp}{\chi_{\perp}}
\newcommand{\spinsuscpar}{\chi_{\parallel}}
\newcommand{\spinsuscparperp}{\chi_{\parallel,\perp}}
\newcommand{\charsusc}{\xi}
\newcommand{\charsuscperp}{\xi_{\perp}}
\newcommand{\charsuscpar}{\xi_{\parallel}}
\newcommand{\charsuscparperp}{\xi_{\parallel,\perp}}

\newcommand{\green}[2]{\langle\langle{#1}\,;{#2}\rangle\rangle}
\newcommand{\D}{\displaystyle}
\newcommand{\Dup}{\displaystyle\rule[-0.5ex]{0mm}{0.5ex}}
\newcommand{\Ddown}{\displaystyle\rule[0.5ex]{0mm}{0.5ex}}

\newcommand{\Ddb}{\displaystyle\rule[2ex]{0mm}{0.5ex}}

\newcommand{\cf}{{\it c.f.\ }}

\newcommand{\oref}{{reference\ }}

\newcommand{\ofig}{{Fig.\ }}
\newcommand{\ofigs}{{Figs.\ }}

\newcommand{\eqref}[1]{{(\ref{#1})}}

\newcommand{\ham}{{\mathcal H}}
\newcommand{\hamat}{{\mathcal H}_a}

\newlength{\incjot}
\title{Response functions of an artificial Anderson atom in the atomic
limit}

\author{Ari T. Alastalo\footnote{Present address:
    VTT Information Technology,
    Tekniikantie 17, Espoo, FIN-02044 VTT, Finland; 
    email: ari.alastalo@vtt.fi}, Markku P. V. Stenberg, and Martti M. Salomaa
\address{Materials Physics
  Laboratory, Helsinki University of Technology, \\ P.O.Box 2200
  (Technical Physics), FIN-02015 HUT, Finland}
}

\runninghead{Ari T. Alastalo, Markku P. V. Stenberg, and Martti
M. Salomaa}{Response functions of an artificial Anderson atom in the
atomic limit}
       
\begin{document}

\begin{abstract}
We consider the spin and pseudospin (charge) response functions of
the exactly soluble Anderson atom model. We demonstrate, in
particular, that a deviation
from the magnetic Curie-law behaviour, appropriate for a free spin
one-half, increases with increasing asymmetry and temperature. In
general, oscillator strength is transferred from the spin degrees of
freedom to the pseudospin modes. We also consider the negative-$U$
Anderson atom and demonstrate that the pseudospin modes are the relevant
low-energy excitations in this case. Especially, the roles of the
spin and charge excitations are interchanged upon reversal of
the intrasite Coulomb repulsion, $U$.

\end{abstract}

\maketitle


\section{INTRODUCTION}
\label{sec:intro}
A single quantum dot behaves like an artificial atom when
electronic confinement in the dot approaches atomic size
\cite{ashoori96,cronenwett,bonadeo}. In such structures, there emerge the
characteristic features of an atomic impurity: the quantization of
charge and energy. It is appropriate to describe these artificial
atoms by means of the Anderson impurity model \cite{meir91,alhassid}.

The Anderson model was first proposed to describe magnetic impurities
in a metal \cite{anderson61}. In the Anderson model, the
nonmagnetic-magnetic transition of the local $d$-state was first
described within the Hartree-Fock (HF) approximation \cite{anderson61}, which
amounts to truncating the model Hamiltonian into a bilinear form in
the fermion operators. This approximation leads to an abrupt phase
transition, whereas the actual change must be a gradual one for a
finite system.  Numerous ingenious approaches to describe the
complicated many-body problem associated with the emergence of the
interacting correlated many-electron problem have been introduced,
such as Green's function methods \cite{rickayzen}, functional-integral
techniques \cite{weller}, numerical renormalization schemes
\cite{krishna12,bulla}, Bethe-Ansatz approaches \cite{wiegmann12}
and noncrossing approximations \cite{riseborough}.
Perturbation theories have been utilized starting both from the
small-$U$ (HF) limit \cite{yamada,yamada2,salomaa12,wei} and from the
atomic ($V_{k}=0$) limit \cite{brinckmann,foglio}.  Furthermore,
interpolation schemes between the HF and atomic limits have been developed
\cite{czycholl,meyer,takagi}.  However, the properties of the atomic
limit, which again is exactly soluble, appears not to have been
thoroughly discussed in the literature \cite{hewson}.

In this paper we want to discuss in detail the coupling of the
correlations which takes place in the magnetic (zero-temperature
susceptibility is divergent) Anderson atom. Relevant information on
the correlations among the electrons may be obtained by considering
the response of a system to an external perturbation.  We derive the
static response functions of the Anderson atom. Our derivation
demonstrates that it is in general important to discuss not only the
spin susceptibilities ($\chi$) but also the charge 
(pseudospin \cite{anderson58}) susceptibilities ($\xi$); the 
superscript zero in $\chi^0$ and $\xi^0$ denotes static response.

We show that the Anderson impurity atom best follows the magnetic Curie law
$\chi^0=(4T)^{-1}$, appropriate for a quantum-mechanical spin
one-half, in the symmetric situation at low temperatures $(T\ll
U)$.  At
high temperatures, on the other hand, the spin and charge degrees of
freedom become equally important and the respective susceptibilities
approach the common limiting value $T\xi^0=T\chi^0\rightarrow 1/8$ for
$T\rightarrow\infty$ in zero field.

In particular, our equations may be applied to negative-$U$
situations.  The negative-$U$ Anderson model was first proposed to
describe the electronic structure of amorphous semiconductors
\cite{anderson75}.  Since then, negative-$U$ behaviour has been
observed, e.g., in the context of high-temperature superconductors
\cite{micnas90}, heavy-fermion systems \cite{taraphder91} and
interstitial defects in semiconductors \cite{harris,ghosh92}.  It has
been observed that in a quantum dot, the second electron in the dot may
be more strongly bound than the first one under certain circumstances
\cite{ashoori92,ashoori93,zhitenev97,ashoori98,brodsky}. 
Essentially two different mechanisms have been
proposed to supply the effective net attraction between the electrons
to cause a quantum dot negative-$U$ properties \cite{wan95,raikh96}.
However, the microscopic origin of the phenomena is still unclear.

The reversal of $U$ changes the roles of spin and charge.  For the
symmetric level configuration and negative $U$, the charge response
functions behave like the spin susceptibilities for positive $U$.
Hence, for $U\le 0$, the spin degrees of freedom are frozen out and
the charge degrees of freedom are the dominating low-energy
excitations.  In what follows, we intend to elucidate the dual roles
played by the magnetic field in connection of spin dynamics and that
of asymmetry in the context of charge dynamics.
 
For high magnetic fields, we find new peak structures in the
longitudinal spin and charge susceptibilities and an associated
threshold behaviour in the corresponding transversal response
functions.  These features are associated with level crossings
occurring between states belonging to different Fock spaces in special
configurations displaying high symmetry.

The rest of this paper is organized as follows. The Anderson
Hamiltonian is presented in Section \ref{sec:hamiltonian}. In Section
\ref{sec:algebra}, we take a close look into the spin and charge
algebras. Section \ref{sec:analytic} is devoted to derivation of
analytic expressions for the spin and charge susceptibilities. In
Section \ref{sec:results}, we present our numerical results. In
Section \ref{sec:discussion} we give
discussion and conclusions.  In Appendix \ref{sec:response} we
collect the central properties of double-time Green's functions that
are utilized throughout the present paper.

\section{MODEL HAMILTONIAN}

\label{sec:hamiltonian}
The Anderson Hamiltonian for magnetic impurities in metals \cite{anderson61}
\begin{eqnarray}
  {\mathcal H} & = &\sum_{k,\sigma}\varepsilon_{k\sigma}n_{k\sigma}+
  \sum_{k,\sigma}\left(V_{k}c_{k\sigma}^{\dagger}d_{\sigma}+
    V_{k}^{\ast}d_{\sigma}^{\dagger}c_{k\sigma}\right)+
  \sum_{\sigma}E_{\sigma}n_{\sigma} \nonumber \\
  & & + Un_{\uparrow}n_{\downarrow}
  \label{AHam}
\end{eqnarray} 
describes the transition of the local $d$-electron orbital from a
nonmagnetic resonant virtual bound state $(\Gamma/U\gg1)$ to a
magnetic atom $(\Gamma/U\ll1)$. Here $U$ is the intra-atomic Coulomb
repulsion energy and $\Gamma=\pi N(0)\langle|V_{k}|^{2}\rangle$ (with
$N(0)$ the density of conduction electron states at the Fermi level,
$V_{k}$ the $d$-level hybridization matrix element and with $\langle
\rangle$ denoting an average over the Fermi surface) is a measure of
the admixture of the local state with energy $E_{\sigma}=E-\sigma B$
(here $\sigma=\pm 1/2$ and $B$ is the external magnetic field).
Furthermore, $c_{k\sigma}^{\dagger}$, $c_{k\sigma}$,
$d_{\sigma}^{\dagger}$ and $d_{\sigma}$ are the creation and
annihilation operators for electrons in the conduction band and in the
impurity state, respectively, and the corresponding occupation-number
operators are $n_{k\sigma}=c_{k\sigma}^{\dagger}c_{k\sigma}$ and
$\nsigb=\dsd\ds$.  The conduction-electron dispersion relation is
denoted with $\varepsilon_{k\sigma}$.  In the nonmagnetic ($d$-spin
susceptibility is finite at $T=0$) $U=0$ limit, the Hamiltonian
(\ref{AHam}) is of bilinear form and hence the static and dynamic
properties of the impurity spin may be obtained exactly in closed form
\cite{salomaa2}. In this paper we discuss the properties of the
Anderson model in the atomic ($V_{k}=0$) limit
\begin{equation}
  \hamat = \Sigma_{\sigma} E_\sigma n_{\sigma} + U\nyna,
  \label{eq:hamatom}
\end{equation} 
for which only the last two terms in the Hamiltonian (\ref{AHam})
remain.

\section{SPIN AND CHARGE ALGEBRAS}
\label{sec:algebra}
Here we form the spin and charge
operators ${\vec{S}}$ and ${\vec{Q}}$. We also summarize some of their
properties.  We define the operator-valued spinor $\psi\equiv\left(
  d_{\uparrow} , d_{\downarrow}\right)^T$, and introduce spin-operator
components as $S_\alpha\equiv\psi^\dagger\sigma_\alpha\psi$ for
$\alpha\in\{x,y,z\}$ and $S^\pm\equiv\left(S_x\pm
  S_y\right)/\sqrt{2}$. Here $\sigma_\alpha$ are the usual Pauli spin
matrices.  Thus we obtain
\begin{eqnarray}
  S^+ &=& \frac{1}{\sqrt{2}} d_{\uparrow}^{\dagger} d_{\downarrow}
  \label{eq:Soperators2} \\
  S^- &=& \frac{1}{\sqrt{2}} d_{\downarrow}^{\dagger} d_{\uparrow}
  \label{eq:Soperators3} \\
  S_z &=& \frac{1}{2}\left( n_{\uparrow}- n_{\downarrow} \right)
  \label{eq:Soperators1} \\
  S_z^2&=&S_x^2=S_y^2=\frac{1}{4}\left( n_{\uparrow} + n_{\downarrow}-
    2 n_{\uparrow} n_{\downarrow} \right)=\frac{1}{3}S^2.
  \label{eq:sz2}
\end{eqnarray}
With the help of the spinor $\phi\equiv\left(d_{\uparrow} ,
  d_{\downarrow}^{\dagger}\right)^T$, the charge-operator components
may be expressed analogously in the form
$Q_\alpha\equiv\phi^\dagger\sigma_\alpha\phi$.  We find

\begin{eqnarray}
  Q^+ &=& \frac{1}{\sqrt{2}} d_{\uparrow}^{\dagger} d_{\downarrow}^{\dagger}
  \label{eq:Qoperators2} \\
  Q^- &=& \frac{1}{\sqrt{2}} d_{\downarrow} d_{\uparrow}
  \label{eq:Qoperators3} \\
  Q_z &=& \frac{1}{2}\left( n_{\uparrow} + n_{\downarrow}-1 \right)
  \label{eq:Qoperators1} \\
  Q_z^2&=&Q_x^2=Q_y^2=\frac{1}{4}\left( 1- n_{\uparrow}-n_{\downarrow}+
    2 n_{\uparrow} n_{\downarrow} \right)
    = \frac{1}{3}Q^2.
  \label{eq:qz2}
\end{eqnarray}
 From the operators $\vec{S}$ and $\vec{Q}$, we can easily
form a spin-1/2 algebra $\vec{C}$ which obeys the canonical spin
commutation relations $\left[C_i,C_j\right]_-=i\epsilon_{ijk}C_k$ with
$C^2=c(c+1)$ for $c=1/2$, as follows.  We note that $\vec{S}$ and
$\vec{Q}$ obey by construction the canonical commutation relations.
Moreover, for the atomic Anderson Hamiltonian we find that $S_z$ and
$Q_z$ and thus $S_z^2$ and $Q_z^2$ (therefore, also $S^2$ and $Q^2$)
are constants of motion.  However, $\langle S^2 \rangle=\frac{3}{4}$
is not obeyed, unless $\langle Q^2 \rangle=0$, and vice versa.
Furthermore, one easily finds that $\left[ S_i,Q_j \right]=0$ for all
$i,j$.  Consequently, a spin-1/2 algebra $\vec{C}$ can be formed as
$\vec{C}\equiv\vec{S}+\vec{Q}$.  One easily sees that
$C^2=S^2+Q^2=\frac{3}{4}$, as required for a spin of fixed magnitude
$1/2$. This clearly demonstrates the importance of considering both
spin and charge degrees of freedom.

\section{EXPRESSIONS FOR THE SPIN AND CHARGE SUSCEPTIBILITIES}
\label{sec:analytic}
By utilizing the Hubbard operators \cite{hubbard}

\begin{eqnarray}
  \As{1} &\equiv& \left( 1-\nmsb \right)\ds
  \label{eq:as2def1}\\
  \As{2} &\equiv& \nmsb\ds
  \label{eq:as2def2} 
\end{eqnarray}
(for a recent discussion of supersymmetric Hubbard operators, see 
 \cite{hopkinson}) or the spin-flip and charge-transfer operators 
\begin{eqnarray}
  \As{3} &\equiv& \dmsd\ds = \sqrt{2}\ S^\pm
  \label{eq:as4def1}\\
  \As{4} &\equiv& \dms\ds = \sqrt{2}\ Q^\pm
  \label{eq:as4def2} 
\end{eqnarray}
the atomic Hamiltonian (\ref{eq:hamatom}) may be
represented as a bilinear form in two alternative pictures 
\begin{eqnarray}
  \hamat &=& \Sigma_\sigma E_\sigma\left(\Asd{1}\As{1} + \Asd{2}\As{2}\right)
  +U\Asd{2}\As{2} \label{eq:hammol21}\\
  &=& \Sigma_\sigma E_\sigma\left(\Asd{3}\As{3} + \Asd{4}\As{4}\right)
  +U\Asd{4}\As{4}. 
  \label{eq:hammol22} 
\end{eqnarray}
Action of the operators $\As{1}$, $\As{2}$, $\As{3}$ and $\As{4}$
and their hermitean adjoints on the states $\st$, $\sty$, $\sta$ and
$\stya$ is shown in Table \ref{operatortable}.  These operators are
also eigenoperators of the Hamiltonian
\begin{equation}
  [\As{k},\hamat] = a_{k\sigma}(E,U,B) \As{k},
  \label{eq:Aeig}
\end{equation}
where $a_{k\sigma}(E,U,B)$ is a scalar-valued function of the model
parameters.  Consequently, it is easy to obtain the following
anticommutator (+) functions directly from the equations of motion
(\ref{eq:greenmotion1}) and (\ref{eq:greenmotion2}) (no
coupling to higher-order Green's functions)

\begin{eqnarray}
  \green{\As{1}}{\Asd{1}}_z^+ &=& \frac{\Dup\langle 1-\nmsb \rangle}{\Ddown z-E_\sigma}
  \label{eq:Agreen1}\\
  \green{\As{2}}{\Asd{2}}_z^+ &=& \frac{\Dup \langle \nmsb \rangle}{\Ddown z-E_\sigma-U}
  \label{eq:Agreen2}\\
  \green{\As{3}}{\Asd{3}}_z^+ &=& \frac{\Dup \langle \nsigb+\nmsb-2\nsigb\nmsb \rangle}{\Ddown z+2\sigma B}
  \label{eq:Agreen3}\\
  \green{\As{4}}{\Asd{4}}_z^+ &=& \frac{\Dup \langle 1-\nsigb-\nmsb+2\nsigb\nmsb \rangle}{\Ddown z-(2E+U)}.
  \label{eq:Agreen4} 
\end{eqnarray}
The corresponding commutator $(-)$ functions are found by
replacing the expectation values in Eqs.
\eqref{eq:Agreen1}-\eqref{eq:Agreen4}, respectively, with
$\langle (1-\nmsb)(1-2\nsigb) \rangle$, $\langle \nmsb-2 \nsigb\nmsb
\rangle$, $\langle \nmsb-\nsigb \rangle$ and $\langle 1-\nsigb-\nmsb
\rangle$.  Note that for the $d$ electron we have $\ds=\As{1}+\As{2}$
and for the $d$-electron propagator we find: $
\green{\ds}{\dsd}_z^+=\green{\As{1}}{\Asd{1}}_z^+ +
\green{\As{2}}{\Asd{2}}_z^+ =\langle 1-n_{-\sigma} \rangle
/\left(z-E_\sigma\right) + \nms /\left( z-E_\sigma-U\right), $ with
poles at the single-particle eigenenergies of the atomic Hamiltonian
(\ref{eq:hamatom}), $z=E_\sigma$ and $z=E_\sigma+U$.
\begin{table}
\caption{\label{operatortable}Operators $\As{1}$, $\As{2}$, $\As{3}$ and $\As{4}$ 
    and their Hermitean adjoints describe all the possible transitions 
    between the states $\st$, $\sty$, $\sta$ and $\stya$. Notation 
    means that one obtains the final state $|f\rangle$ through the operation 
    of an entry in the table on the initial state $|i\rangle$.}
  \center
  \vspace{\baselineskip}
  $\begin{array}{c||c} 
    \begin{array}{cc} & \mbox{initial} \\
      \mbox{final} & \end{array}
    & 
    \begin{array}{cccc} \multicolumn{4}{c}{|i\rangle} \\
      \st & \sty & \sta & \stya \end{array}                         
    \\[2ex] \hline\hline
    \renewcommand{\arraystretch}{1.4}
    \begin{array}{cc}  
      |f\rangle & \begin{array}{c} \st \\ \sty \\ \sta \\ \stya \end{array}
    \end{array}
    & 
    \renewcommand{\arraystretch}{1.4}
    \begin{array}{cccc}
      \uni    & \Ay{1}  & \Aa{1}  & \Ay{4} \\
      \Ayd{1} & \uni    & \Ayd{3} & \Aa{2} \\
      \Aad{1} & \Ay{3}  & \uni    & \Ay{2} \\
      \Ayd{4} & \Aad{2} & \Ayd{2} & \uni
    \end{array}
  \end{array}$
\end{table}

Considering the time correlation functions (\ref{eq:corr1}) and
(\ref{eq:corr2}) for the Green's functions
(\ref{eq:Agreen1})--(\ref{eq:Agreen4}), one finds after some algebra

\begin{equation}
  \nsig=\frac{\D\nsignms}{\D f(E_{-\sigma}+U)},
  \label{eq:nsres}
\end{equation}
where
\begin{equation}
  \nsignms = \frac{1-f(E_\sigma)-f(E_{-\sigma})}
  {\D\frac{1}{f(2E+U)} - \frac{f(E_\sigma)}{f(E_\sigma + U)} - 
    \frac{f(E_{-\sigma})}{f(E_{-\sigma} + U)}}
  \label{eq:nynaob}
\end{equation}
is the correlated double occupancy and $f$ is the Fermi function.

Now we turn our attention to the longitudinal and transversal spin and
charge response functions of the Anderson atom. They may be defined,
for complex frequencies, as

\begin{eqnarray}
  \spinsuscperp(z)&\equiv&-\green{S^+}{S^-}_z^- =-\frac{1}{2}\green{\Aa{3}}{\Aad{3}}_z^- 
  \label{eq:spinsuscperp}\\
  \charsuscperp(z)&\equiv&-\green{Q^+}{Q^-}_z^- =-\frac{1}{2}\green{\Ayd{4}}{\Ay{4}}_z^-
  \label{eq:charsuscperp} \\
  \spinsuscpar(z)&\equiv&-\green{\Bigl(S_z-\langle S_z \rangle\Bigr)}{\Bigl(S_z-\langle S_z \rangle\Bigr)}_z^-
  \label{eq:spinsuscpar}\\
  \charsuscpar(z)&\equiv&-\green{\Bigl(Q_z-\langle Q_z \rangle\Bigr)}{\Bigl(Q_z-\langle Q_z \rangle\Bigr)}_z^-.
  \label{eq:charsuscpar}
\end{eqnarray}
 The expectation values $\langle S^\pm \rangle$ and
$\langle Q^\pm \rangle$ vanish for $\hamat$ and are thus not needed in
the defining equations \eqref{eq:spinsuscperp} and \eqref{eq:charsuscperp}.
However, $\langle S_z \rangle$ and $\langle Q_z \rangle$ in Eqs.
\eqref{eq:spinsuscpar} and \eqref{eq:charsuscpar} are generally
nonzero.  In connection with the $O(3)$-symmetric Anderson model and
the two-channel Kondo model, the charge operators are called isospin
operators \cite{bradley} and the parallel charge susceptibility is
defined as a $Q_z-Q_z$ response as above but without extracting the
mean.

The transversal spin susceptibility ($(-)$ function) is readily found
from Eq. \eqref{eq:Agreen3} (changing the expectation value as
described in the text after \eqref{eq:Agreen3})

\begin{equation}
  \spinsuscperp (z) = -\frac{\langle S_z \rangle}{z-B}.
  \label{eq:dynTX}
\end{equation}
Consequently, the familiar static zero-frequency limit of
$\spinsuscperp (z)$ is found with \eqref{eq:hilbert}
\begin{equation}
  \spinsuscperp^{0}=\int_{-\infty}^\infty\!\frac{d\omega}{\pi}\frac{ \spinsuscperp^{''}(\omega) }
  {\omega}=\frac{\langle S_z \rangle}{B}.
  \label{eq:statTX0}
\end{equation}
Similarly, for the transversal charge susceptibility one obtains
\begin{equation}
  \charsuscperp (z) = -\frac{\langle Q_z \rangle}{z+(2E+U)}
  \label{eq:dynTK}
\end{equation}
and
\begin{equation}
  \charsuscperp^{0}=\int_{-\infty}^\infty\!\frac{d\omega}{\pi}\frac{ \charsuscperp^{''}(\omega) }
  {\omega}=-\frac{\langle Q_z \rangle}{2E+U}.
  \label{eq:statTK0}
\end{equation}
Here $2E+U$ measures the asymmetry of the level configuration with
respect to the Fermi level, vanishing in the symmetric ($E=-U/2$)
situation.  Thus asymmetry behaves for the charge degrees of freedom
as the magnetic field for the spin degrees of freedom (compare 
Eqs. \eqref{eq:statTX0} and \eqref{eq:statTK0}).

For the parallel spin response in Eq. \eqref{eq:spinsuscpar}, one
cannot directly calculate the commutator $(-)$ function with the
equations of motion \eqref{eq:greenmotion1} and
\eqref{eq:greenmotion2} since $[S_z-\langle S_z \rangle,S_z-\langle
S_z \rangle]=0$, and also $[S_z-\langle S_z \rangle,\hamat]=0$.
However the corresponding anticommutator function is easy to find
since $\{S_z-\langle S_z \rangle,S_z-\langle S_z \rangle\}
=2\left(\langle S_z^2 \rangle - \langle S_z \rangle ^2\right)$, after
which the static parallel spin susceptibility $\spinsuscpar^0$ is
obtained with the help of the fluctuation-dissipation theorem
\eqref{eq:flucdis}. The result is
\begin{equation}
  \spinsuscpar^0=\frac{\D\langle S_z^2 \rangle
    - \langle S_z \rangle ^2}{\D T}.
  \label{eq:statTXII0} 
\end{equation}
For the parallel charge response, a similar calculation yields
\begin{equation}
  \charsuscpar^0=\frac{\D \langle Q_z^2 \rangle
    - \langle Q_z \rangle ^2}{\D T}.
  \label{eq:statTKII0}
\end{equation}
The parallel susceptibilities may, furthermore, be calculated from
\begin{equation}
  \spinsuscpar^0=\frac{\D\partial\langle S_z \rangle}{\D\partial B},
  \label{eq:statTXII1}
\end{equation}
and
\begin{equation}
  \charsuscpar^0=-\left.\frac{\partial\langle Q_z \rangle}{\partial (2E+U)}
  \right|_U\,,
  \label{eq:statTKII1}
\end{equation}
where the partial derivation is performed such that the intra-atomic
Coulomb repulsion $U$ is held constant.

The expectation values $\langle S_z \rangle$, $\langle S_z^2 \rangle$,
$\langle Q_z \rangle$ and $\langle Q_z^2 \rangle$ in the above results
for the response functions are expressible through $\nsig$ and
$\nsignms$ as shown in Section \ref{sec:algebra}. Furthermore, since
we know $\nsig$ and $\nsignms$ from Eqs. \eqref{eq:nsres} and
\eqref{eq:nynaob}, respectively, it is now straightforward to find the
susceptibilities.  In terms of the dimensionless parameters
\begin{eqnarray}
    x&\equiv & T/U, \quad y\equiv E/U, \quad b\equiv B/T, \nonumber\\
    {\cal{E}}&\equiv & e^{y/x} + e^{b/2} + e^{-b/2} + e^{(-1-y)/x}
  \label{eq:param}
\end{eqnarray}
the final results are
\label{eq:finres}

\begin{eqnarray}
  T\chi_\perp^0&=&
  \left.{\D\sinh{\!\!\left(\frac{b}{2}\right)}}
  \right/\left({\D b\,{\cal{E}}}\right)
  \label{eq:furtspinsusc} \\
  T\chi_\parallel^0&=&
  \left.\left\{{\D 1 + \frac{1}{2}\cosh{\!\!\left(\frac{b}{2}\right)}
    \Bigl[ e^{y/x} + e^{(-1-y)/x} \Bigr]}
  \right\}\right/{\Ddb {\cal{E}}^2} 
  \label{eq:chiparnum} \\
  T\xi_\perp^0&=&
  \left.{\D x\Bigl[ e^{y/x} - e^{(-1-y)/x} \Bigr]}
  \right/\left[{\D 2\,(1+2y)\,{\cal{E}}}\right]
  \label{eq:furtcharsusc} \\
  T\xi_\parallel^0&=&
  \left.\left\{{\D e^{-1/x} + \frac{1}{2}\cosh{\!\!\left(\frac{b}{2}\right)}
    \Bigl[ e^{y/x} + e^{(-1-y)/x} \Bigr]}
  \right\}\right/{\Ddb {\cal{E}}^2}.
 \nonumber \\
  \label{eq:ksiparnum}
\end{eqnarray}

In what follows, we omit the superscript zero from
$\chi_{\perp,\parallel}^0$ and $\xi_{\perp,\parallel}^0$ since in this
paper we only consider the static response functions.

In addition, we mention the following properties obeyed in the
zero-field and symmetric limits for the static responses

\begin{eqnarray}
  \ \lim_{B\rightarrow 0}\ \ \spinsuscpar&=\ \lim_{B\rightarrow 0}\ \ \spinsuscperp
  \label{eq:spinlim}\\
  \lim_{E\rightarrow -U/2}\charsuscpar&=\lim_{E\rightarrow -U/2}\charsuscperp.
  \label{eq:charlim} 
\end{eqnarray}
 The equalities (\ref{eq:spinlim}) and (\ref{eq:charlim})
again demonstrate that the asymmetry is analogous to an external
magnetic field when one considers the charge response functions
instead of the spin susceptibilities. All the response functions are
symmetric with respect to reversal of the external field
($B\rightarrow-B$).  What is more interesting, however, is that the
susceptibilities are also symmetric with respect to reversal of
asymmetry ($a\equiv 2E+U\rightarrow-a$).

\section{RESULTS}
\label{sec:results}
Here we consider the magnetic-field and level-asymmetry
dependencies of the spin and charge response functions,
$\spinsuscparperp (T)$ and $\charsuscparperp (T)$.  We keep $U$
constant and vary $E$ (or $y$), such that there is a one-to-one
relationship between $y\equiv E/U$ and asymmetry $a\equiv 2E+U=U(2y+1)$.
Consequently, the curves are seen to be drawn for varying level
asymmetries.  Figures \ref{Fig1}--\ref{Fig7} are for $U>0$ while
\ofigs \ref{Fig8}--\ref{Fig12} describe the negative-$U$ situation.
The dashed curves always denote $y<0$, while the solid ones are for
$y\geq 0$. Furthermore, we have labelled the curves such that the
field parameter $b\equiv B/T$ is shown inside the parentheses. When 
the field
parameter is omitted, $b=0$ is implied.  Owing to the above-mentioned
symmetry of the response functions with respect to reversal of
asymmetry, it is convenient to consider positive asymmetries
only for $U>0$ and negative asymmetries for $U<0$.  In particular,
this means that $y\geq-1/2$ below.

The number of illustrations that follow is large. However, we
explore the four response functions to illustrate their
interrelationships and symmetries upon reversal of
$U$ with varying level asymmetries and external field strengths.

\subsection{Positive $U$}
\subsubsection{Zero External Field}
\begin{figure}
  \begin{center}
    \psfig{file=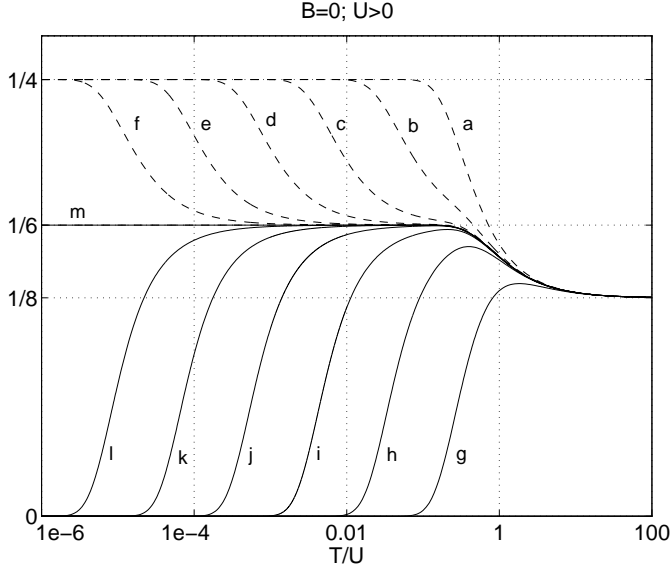,width=0.7\columnwidth}
  \end{center}
\caption{\label{Fig1}Curie parameter $T\chi$ in zero external field 
  $(\chi_{\,\parallel}\equiv\chi_{\perp})$ for positive $U$ as a
  function of level asymmetry.  Here
  $|y|\in\{0\}\bigcup\,\{\frac{1}{2^p}\,|\,p\in\{1,4,7,10,13,16\}\}$,
  such that for the curves "a"\,$\rightarrow$"f"\,
  ("g"\,$\rightarrow$"l"\,)\,,\, $y$ is negative (positive). The curve
  labeled "m"\, denotes $y=0$.}
\end{figure}

\begin{figure}
  \begin{center}
    \psfig{file=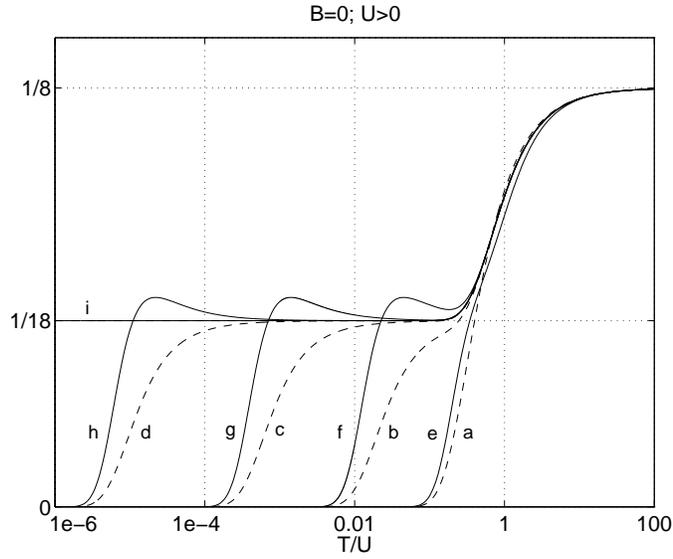,width=0.7\columnwidth}
  \end{center}
\caption{\label{Fig2}Curie parameter for the charge susceptibility
  $T\xi_{\parallel}$ with $U>0$ for vanishing magnetic field as a
  function of asymmetry.  Here
  $|y|\in\{0\}\bigcup\,\{\frac{1}{2^p}\,|\,p\in\{1,5,10,16\}\}$, such
  that for the curves "a"\,$\rightarrow$"d"\, ("e"\,$\rightarrow$"h"\,)\,,\, $y$
  is negative (positive). The curve
  labeled "i"\, denotes $y=0$.}
\end{figure}

Figure \ref{Fig1} shows the spin susceptibility in zero external field
($\spinsuscperp=\spinsuscpar$) for various asymmetries. This is also
given in \oref\cite{krishna12}.  We observe that the Curie
law for a quantum-mechanical spin one half, $T\spinsusc=1/4$, is best
obeyed in the low-temperature limit for the symmetric situation (curve
"a"\,).  However, for $y<0$, the spin susceptibility always finally
rises to the level $(4T)^{-1}$ for low enough temperatures. This
happens simultaneously with the depression of the parallel charge
susceptibility, see \ofig\ref{Fig2}.  Also the perpendicular charge
susceptibility vanishes at low temperatures for all configurations,
see \ofig\ref{Fig7}a.  It is thus legitimate to state that the spin
degrees of freedom are the relevant low-energy (or strong-$U$)
excitations for $U>0$.  Furthermore, we find
for zero field ($\langle S_z \rangle=0$)
\begin{equation}
  T\chi \!+\! T\xi_\parallel \!+\! \langle Q_z \rangle^2 =
  T\chi \!+\! T\xi_\parallel \!+\! \left( \frac{2y+1}{T/U} \right)^2 
  \left( T\xi_\perp \right)^2=\frac{1}{4}.
  \label{eq:TplusK}
\end{equation}
Therefore, as the charge susceptibilities vanish for low temperatures,
we obtain $T\spinsusc=1/4$, as already stated.  For high temperatures
($T\gg U$), on the other hand, the oscillator strength becomes evenly
distributed among the spin and charge degrees of freedom and the spin
and charge susceptibilities approach the common value
$\lim_{T/U\rightarrow\infty}T\spinsusc=\lim_{T/U\rightarrow\infty}T\charsusc=1/8$
in zero field for all values of $y$, \cf\ofigs\ref{Fig1}, \ref{Fig2}
and \ref{Fig7}a.


The $y=0$ configuration is a special case.  From the general result
for the spin-state occupations in Eq. (\ref{eq:nsres}), it is easy to
see that here the occupation numbers satisfy: $\nup=\ndown=1/3$, which
yields $\langle Q_z \rangle^2=1/36$, see Eq. (\ref{eq:qz2}).
Consequently, one finds that the low-temperature limits in zero field
for $y=0$: $T\spinsusc= 1/6$, $T\charsuscpar= 1/18$ and
$T\charsuscperp\rightarrow 0$ are consistent with the result in Eq.
(\ref{eq:TplusK}).

\subsubsection{Finite Field}
\begin{figure}
\hfill
\begin{minipage}[t]{.45\textwidth}
  \begin{center}
    \hspace{0.1\columnwidth}
\psfig{file=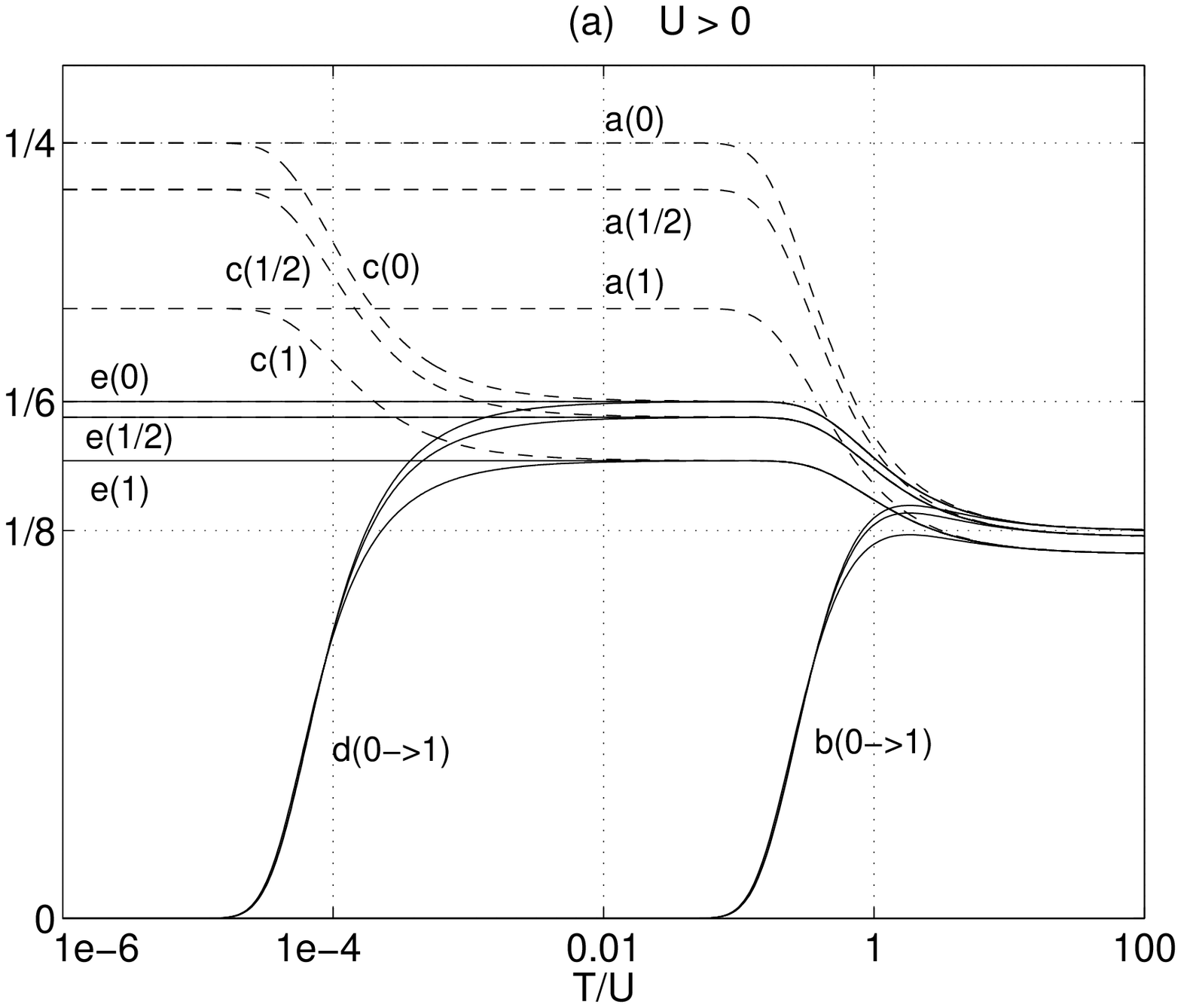,width=\columnwidth}
\end{center}
\end{minipage}
\hfill
\begin{minipage}[t]{.45\textwidth}
  \begin{center}
			
    ~\psfig{file=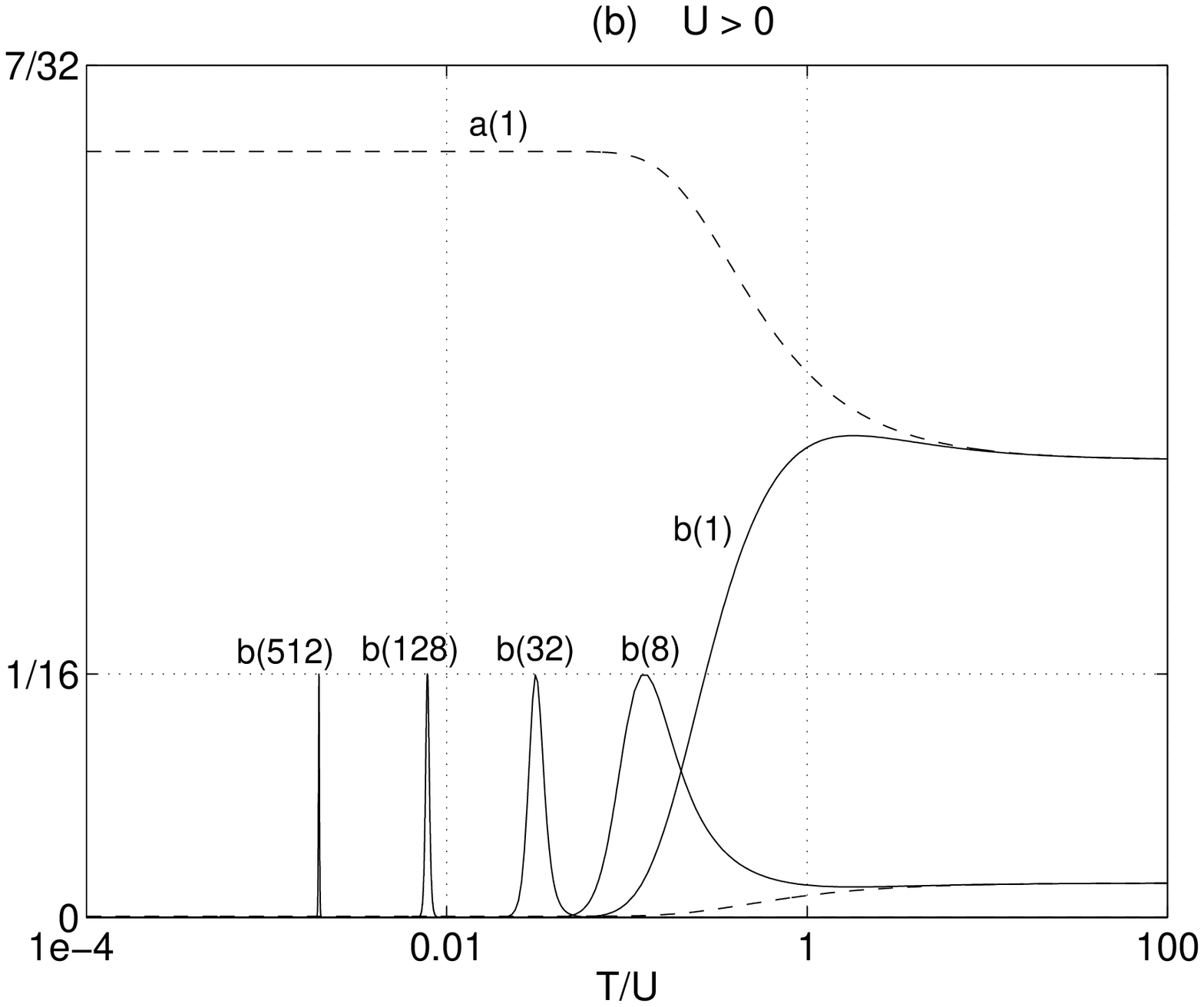,width=\columnwidth}
  \end{center}
\end{minipage}
\hfill
\caption{\label{Fig3}Magnetic-field dependence of the 
  longitudinal magnetic susceptibility $\chi_{\parallel}(B,T)$ in (a)
  moderate and (b) extreme fields. The curves "a"\, and "b"\, are
  for $|y|=1/2$ while "c"\, and "d"\, denote $|y|=1/(2^{13})$. The curves
  for which $y=0$ are labeled "e"\,.  Magnetic field is measured in
  terms of $b=B/T$ which is indicated in the parentheses in this figure.
}
\end{figure}

\begin{figure}
\hfill
\begin{minipage}[t]{.45\textwidth}
  \begin{center}
    \hspace{0.1\columnwidth}
\psfig{file=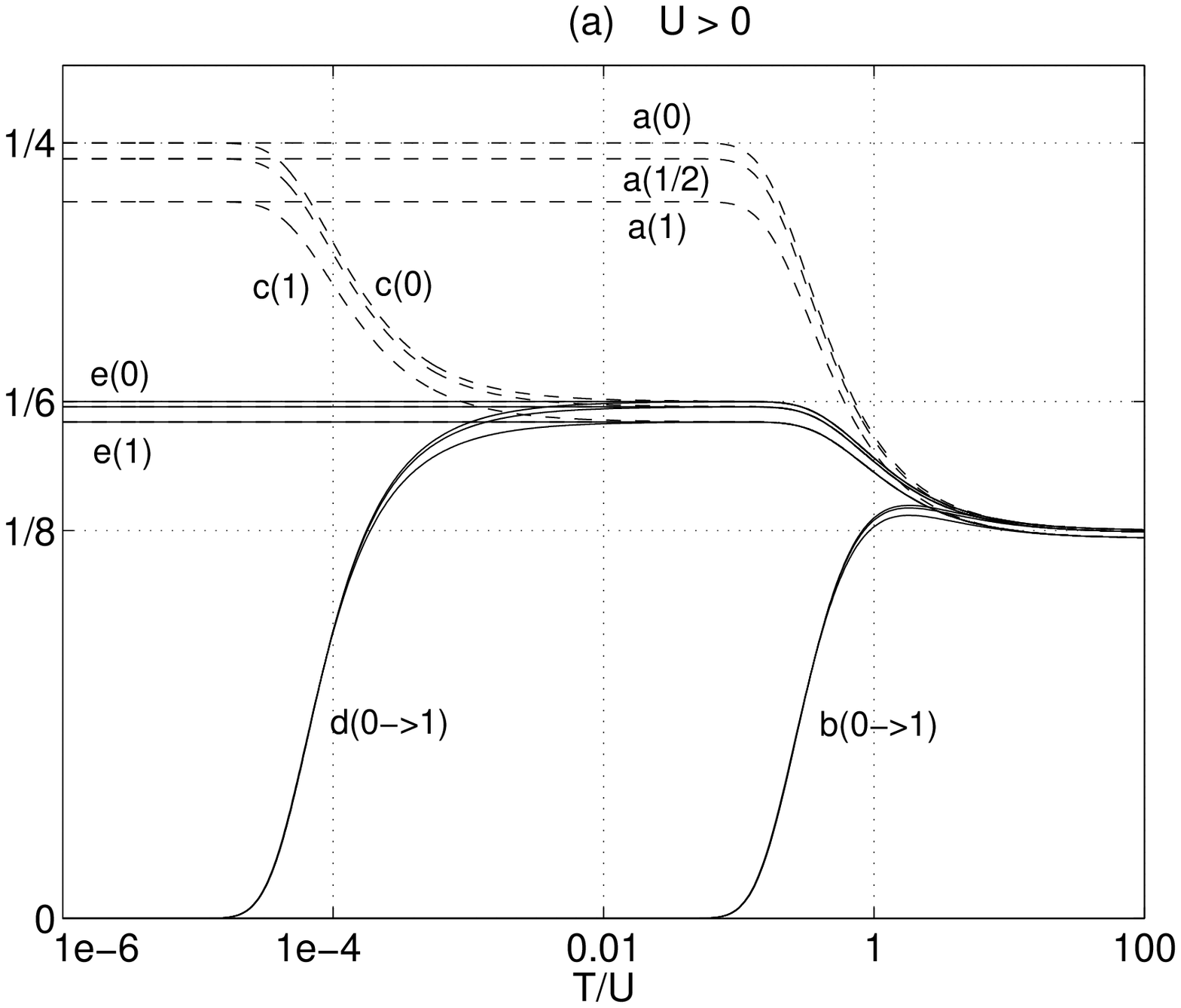,width=\columnwidth}
\end{center}
\end{minipage}
\hfill
\begin{minipage}[t]{.45\textwidth}
  \begin{center}
    ~\psfig{file=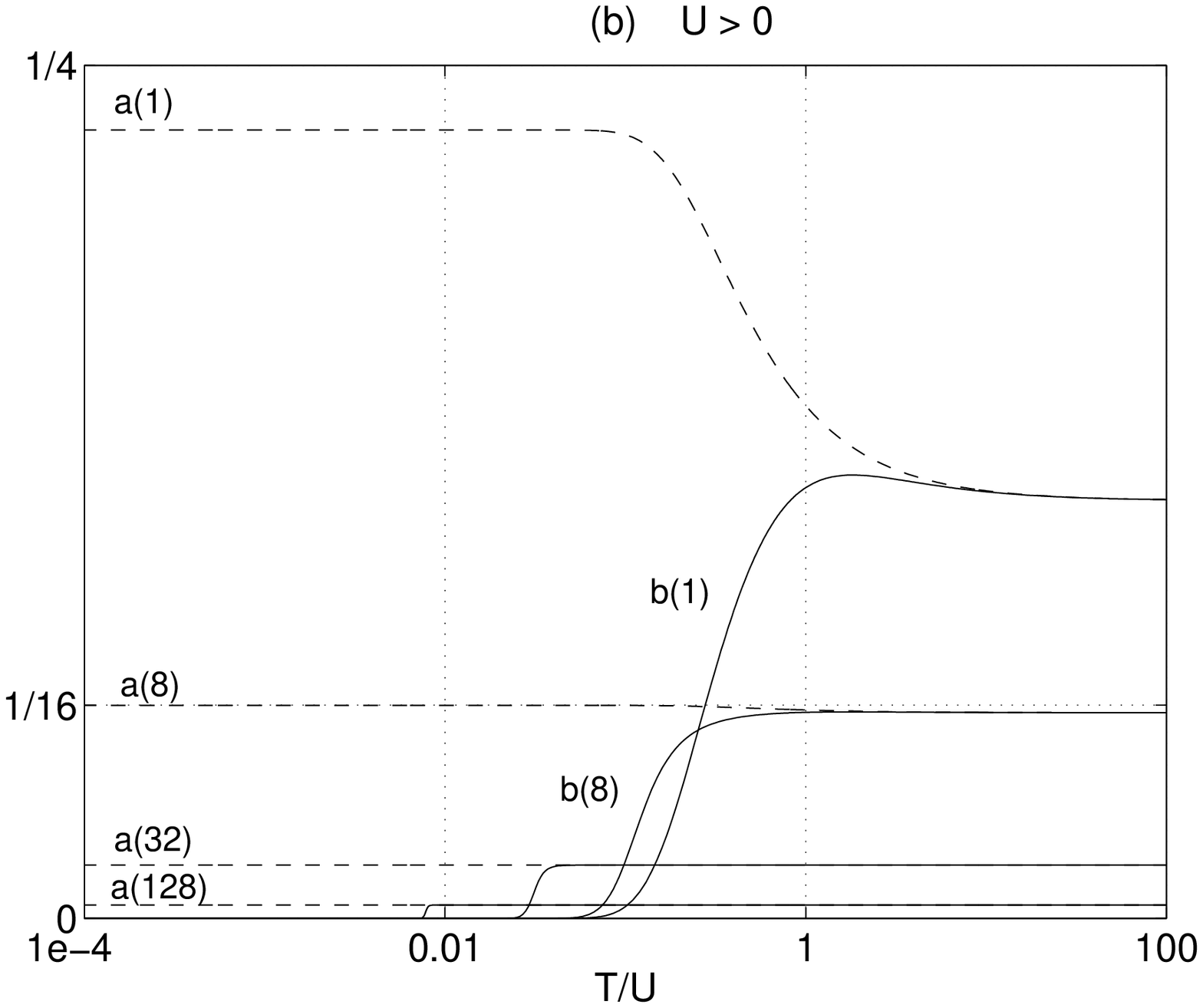,width=\columnwidth}
  \end{center}
\end{minipage}
\hfill
\caption{\label{Fig4}Magnetic field dependence of the 
  transversal spin susceptibility $\chi_{\perp}(B,T)$ in (a) moderate
  and (b) extreme fields. }
\end{figure}

\begin{figure}
  \begin{center}
    \psfig{file=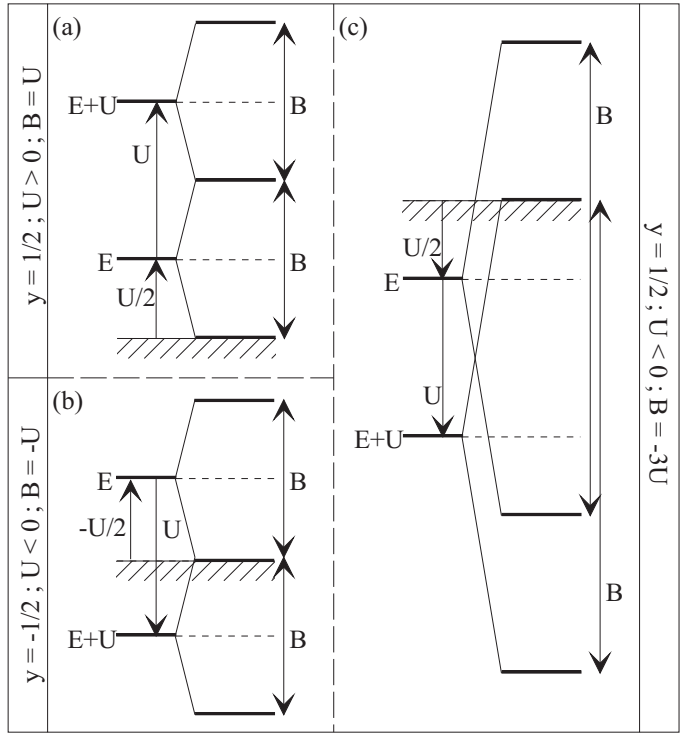,width=0.7\columnwidth}
  \end{center}
\caption{\label{Fig5}Special values of large applied magnetic 
  fields yield level configurations with particular symmetry. These
  level schemes correspond to distinct features in the magnetic and
  charge response functions (a: c.f., \ofigs\ref{Fig3}b, \ref{Fig4}b,
  \ref{Fig6}b, \ref{Fig7}b; b: c.f., \ofigs\ref{Fig11}b,
  \ref{Fig12}b; c: \cf, \ofigs\ref{Fig9}b, \ref{Fig10}b,
  \ref{Fig11}b, \ref{Fig12}b). }
\end{figure}

Figures \ref{Fig3}a and \ref{Fig4}a show the field dependencies of the
longitudinal and perpendicular spin susceptibilities, respectively,
for weak fields. One observes that the longitudinal component is more
strongly affected by the external magnetic field.  Furthermore, in
high fields -- \ofigs\ref{Fig3}b and \ref{Fig4}b -- the longitudinal
spin response displays peaks of invariant height for $y>0$, whereas
the transversal function exhibits threshold behaviour. These peaks and
thresholds correspond to the situation where one of the localized
energy levels crosses the Fermi level as shown in \ofig\ref{Fig5}a for
$y=1/2$.  The height of the peaks in $T\spinsuscpar$ is $1/16$,
irrespective of the absolute value of $y$.

\begin{figure}
\hfill
\begin{minipage}[t]{.45\textwidth} 
  \begin{center}
    \hspace{0.1\columnwidth}\psfig{file=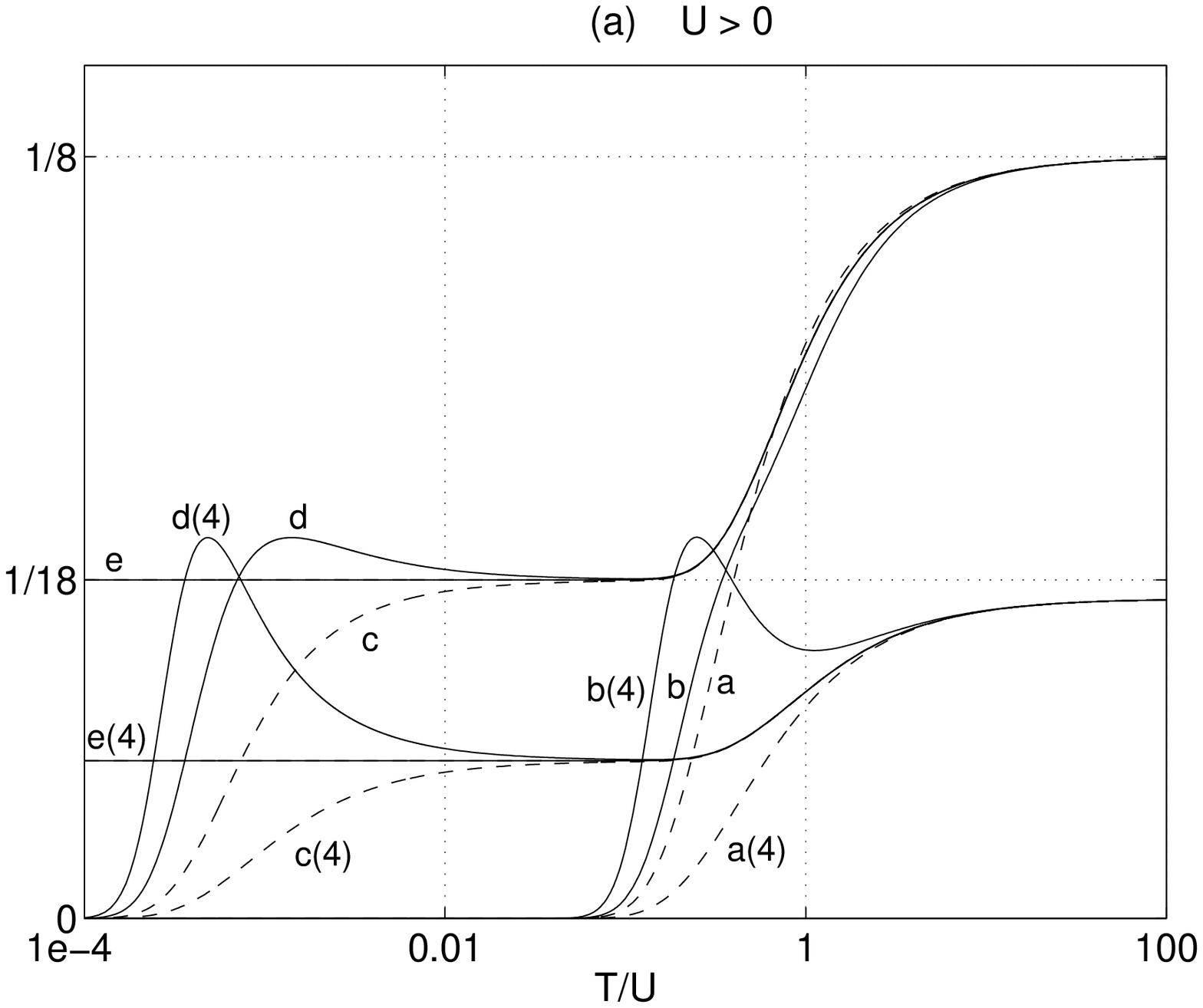,width=\columnwidth}
   \end{center}
\end{minipage}
\hfill
\begin{minipage}[t]{.45\textwidth}
\begin{center}
  \hspace{0.1\columnwidth}\psfig{file=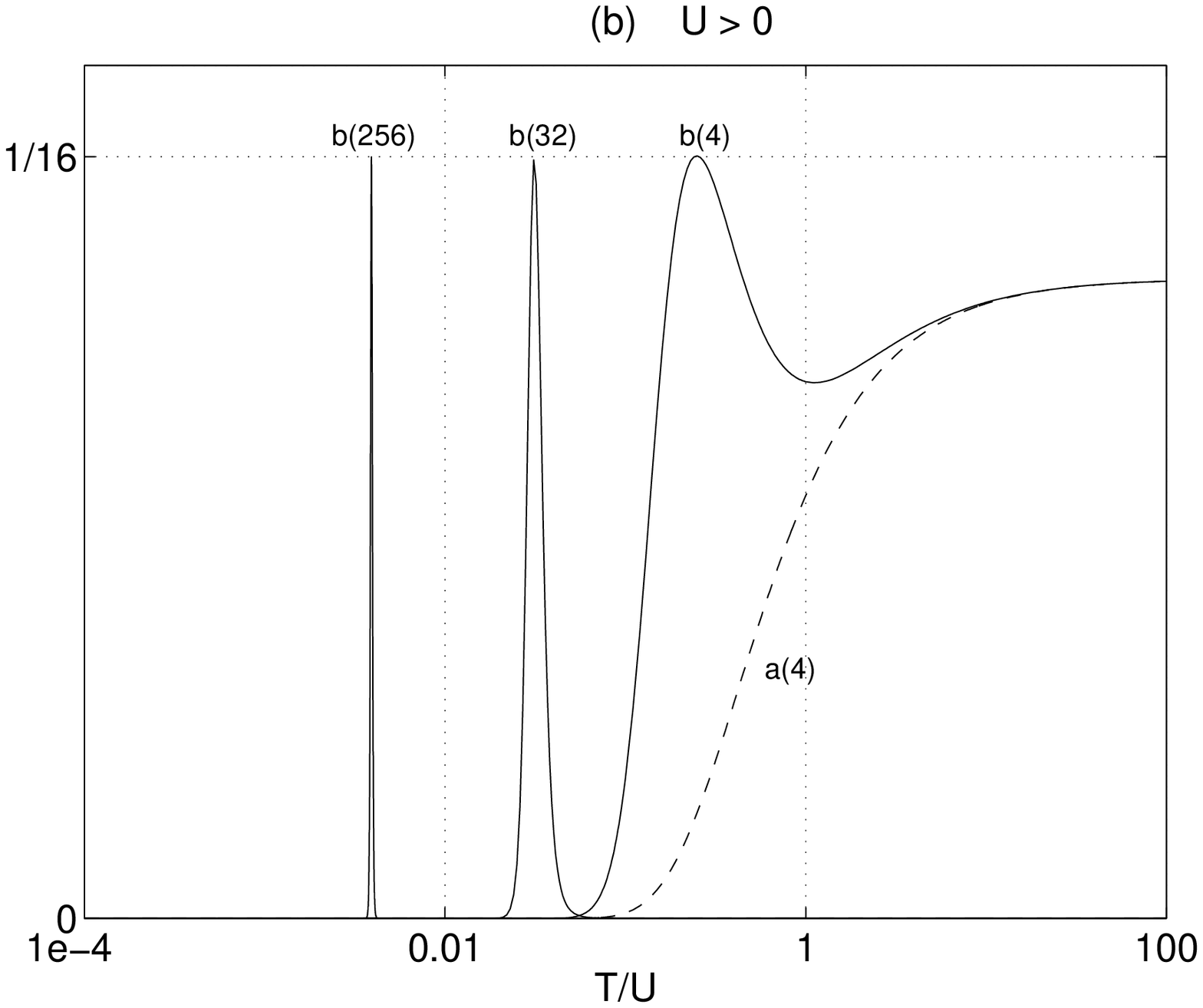,width=\columnwidth}
  \end{center}
\end{minipage}
\hfill
\caption{\label{Fig6}Field dependence of the parallel charge 
  susceptibility $\xi_{\parallel}$ in (a) moderate and (b)
  extreme fields.  The curves "a"\, and "b"\, are for $|y|=1/2$ while "c"\,
  and "d"\, denote $|y|=1/(2^{10})$. The curves for which $y=0$ are
  labeled "e"\,.  The zero-field situation was already considered in
  \ofig\ref{Fig2}.  }
\end{figure}

\begin{figure}
\hfill
\begin{minipage}[t]{.45\textwidth} 
  \begin{center}
    \hspace{0.1\columnwidth}\psfig{file=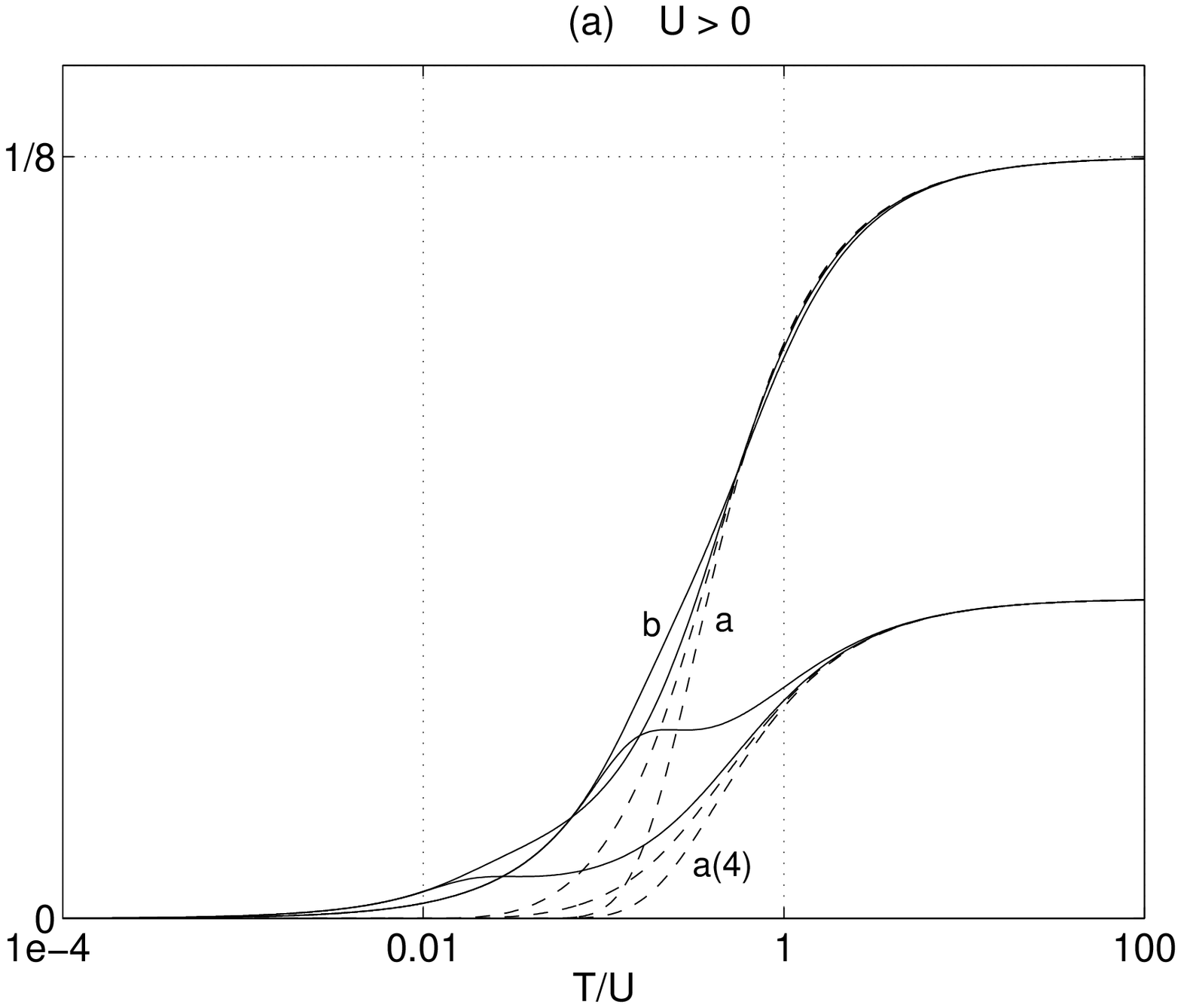,width=\columnwidth}
\end{center}
\end{minipage}
\hfill
\begin{minipage}[t]{.45\textwidth}
\begin{center}
    ~\psfig{file=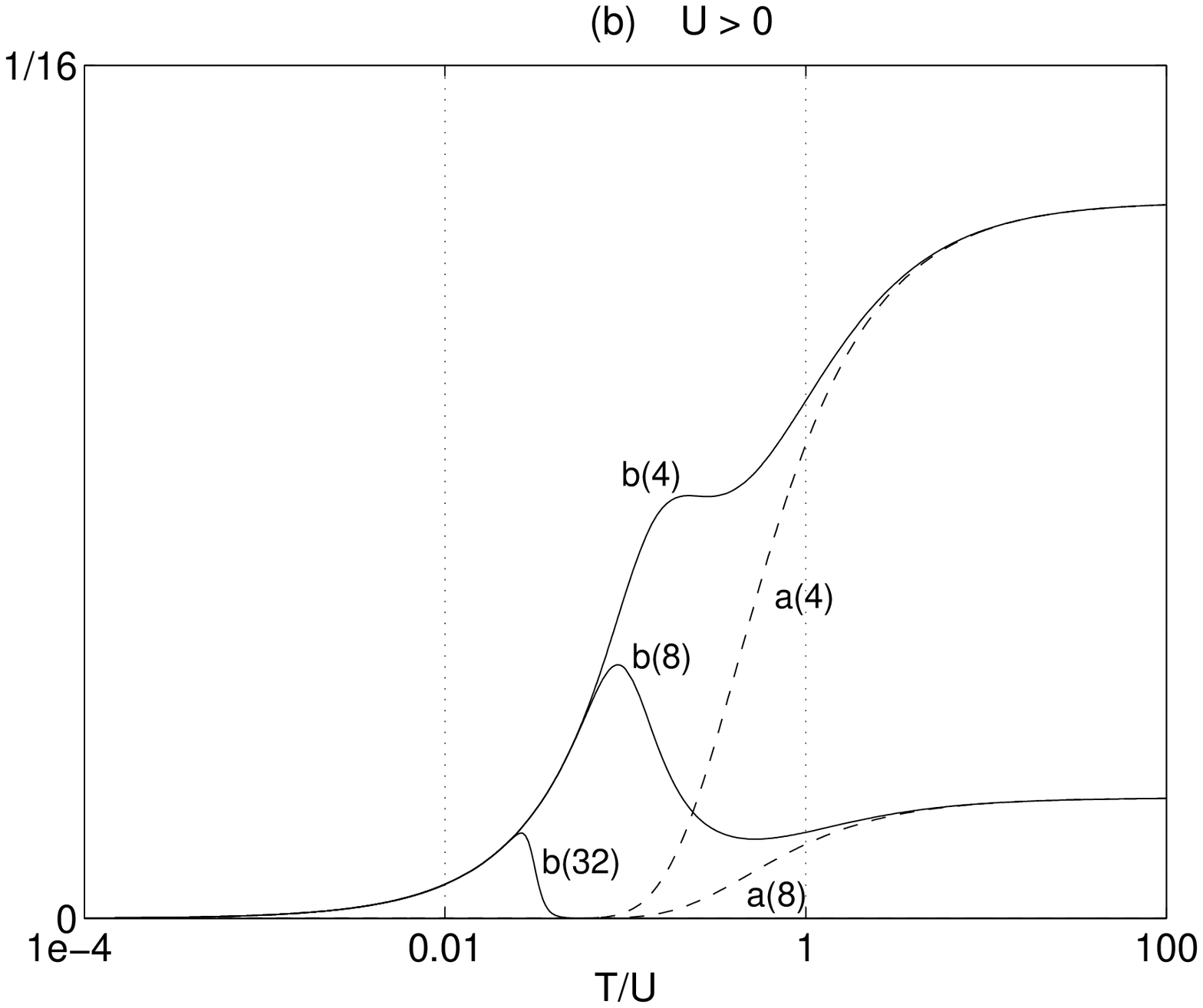,width=\columnwidth}
  \end{center}
\end{minipage}
\hfill
\caption{\label{Fig7}Field- and asymmetry dependence of the 
  perpendicular charge susceptibility $\xi_{\perp}$ for $U>0$ in (a)
  moderate and (b) extreme fields.  The curves "a"\, and "b"\, are for
  $|y|=1/2$ while "c"\, and "d"\, denote $|y|=1/16$.}
\end{figure}

The parallel and perpendicular charge response functions for weak and
strong fields are considered in \ofigs\ref{Fig6} and \ref{Fig7}. Also
for the charge susceptibility, the parallel component is peaked at
high fields for $y>0$ at the level $1/16$.  This is shown in
\ofig\ref{Fig6}b for $y=1/2$ in which case the peak again occurs in
the level configuration of \ofig\ref{Fig5}a.  Simultaneously, the
perpendicular charge susceptibility shows threshold dependence, see
\ofig\ref{Fig7}b. In \ofig\ref{Fig7}a, for $T\charsuscperp$ the curve
for $y=0$ would lie between the $|y|=1/16$ curves shown in the figure.

\subsection{Negative $U$}
\begin{figure}
  \begin{center}
    \psfig{file=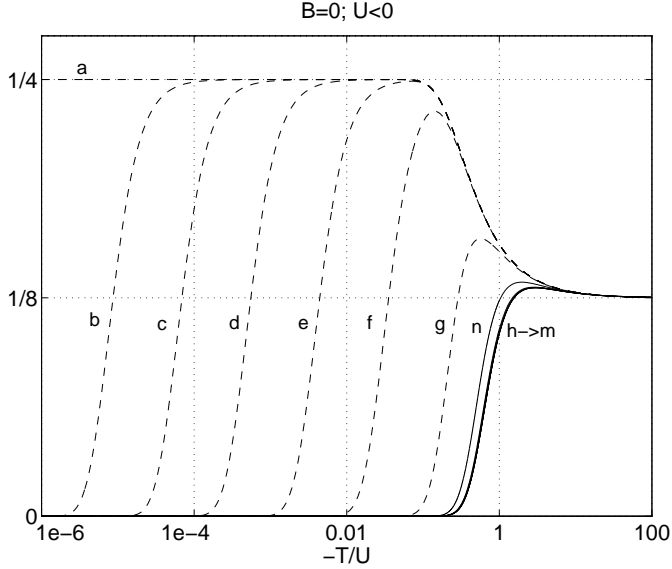,width=0.7\columnwidth}
  \end{center}
\caption{\label{Fig8}Asymmetry dependence of the longitudinal 
  zero-field charge susceptibility, $\xi_{\,\parallel}(B=0,T)$, for
  $U<0$.  Here
  $|y|\in\{\frac{1}{2}\}\bigcup\,\{\frac{2^p-1}{2^{p+1}}\,|\,p\in\{16,13,10,7,4,1\}\}$,
  such that for the curves "a"\,$\rightarrow$"g"\, ("h"\,$\rightarrow$"n"\,),
  $y$ is negative (positive).}
\end{figure}

\begin{figure}
\begin{minipage}[t]{.45\textwidth}
  \begin{center}
    \hspace{0.1\columnwidth}\psfig{file=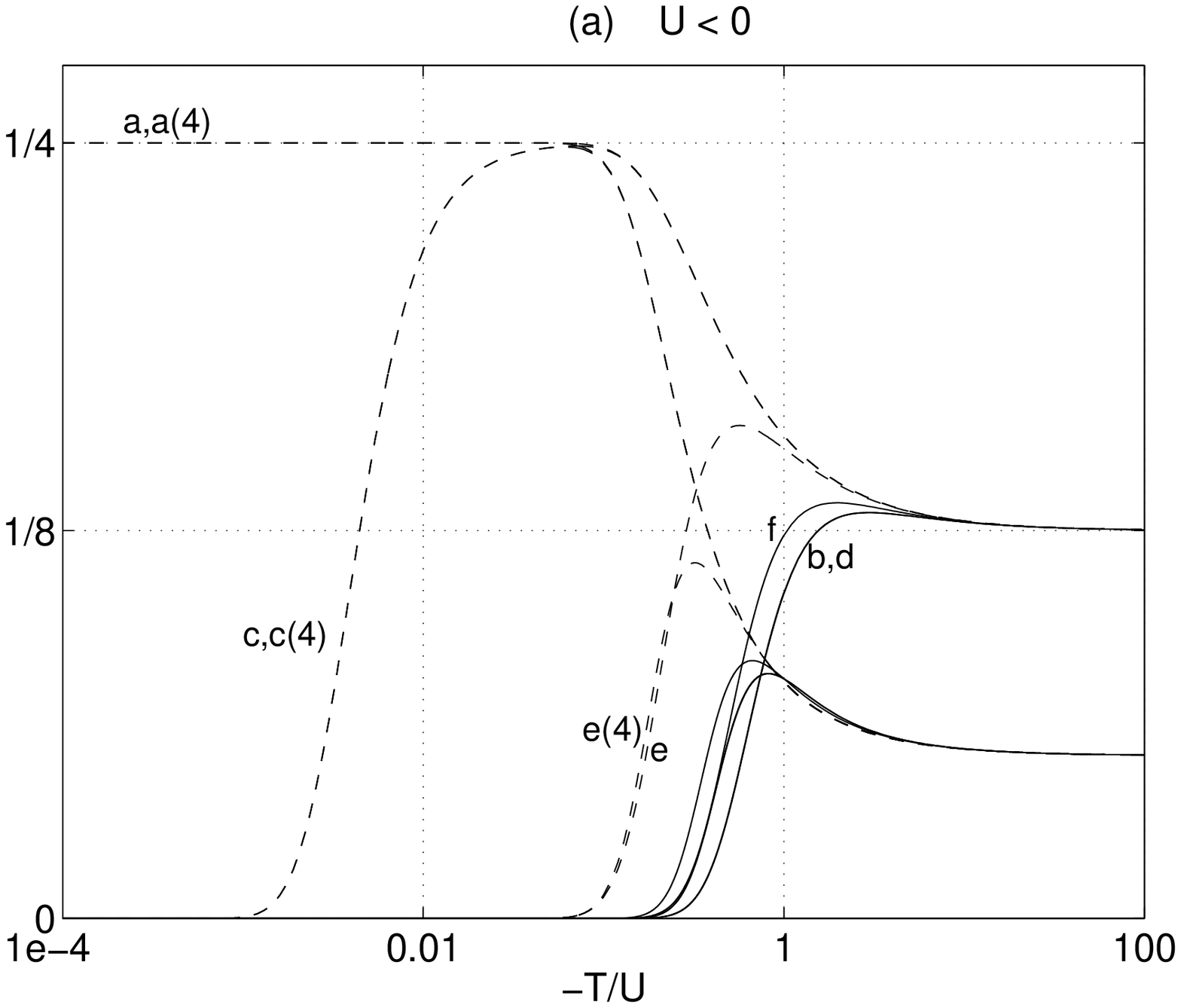,width=\columnwidth}
\end{center}
\end{minipage}
\hfill
\begin{minipage}[t]{.45\textwidth}
\begin{center}
    ~\psfig{file=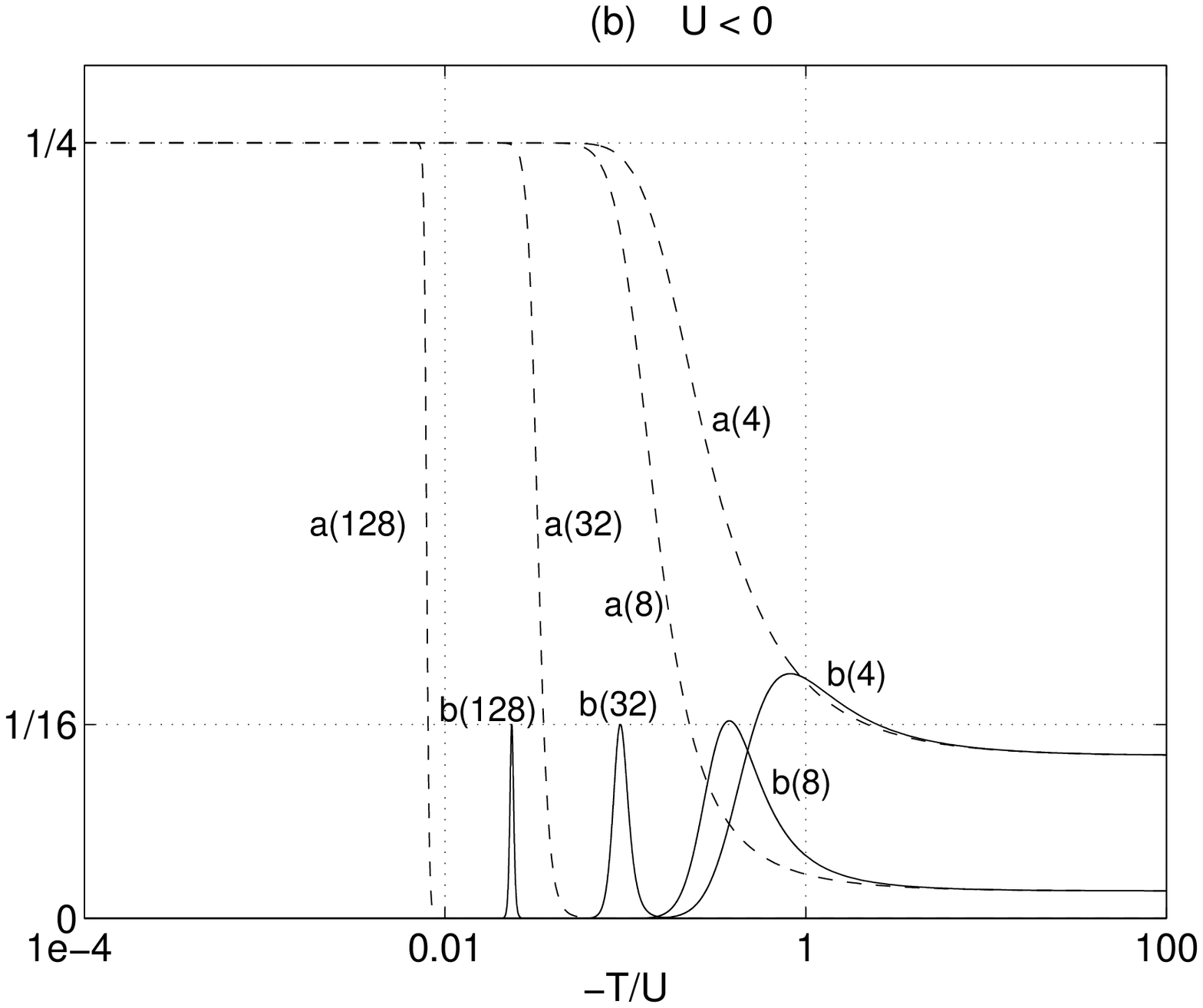,width=\columnwidth}
  \end{center}
\end{minipage}
\caption{\label{Fig9}Field dependence of $\xi_{\parallel}$, for $U<0$
  in (a) moderate and (b) extreme fields.  The curves "a"\, and "b"\, are
  for $|y|=1/2$, "c"\, and "d"\, denote $|y|=127/256$ and the curves for
  which $|y|=1/4$ are labeled "e"\, and "f"\,. }
\end{figure}

\begin{figure}
\hfill
\begin{minipage}[t]{.45\textwidth}
  \begin{center}
    \hspace{0.1\columnwidth}\psfig{file=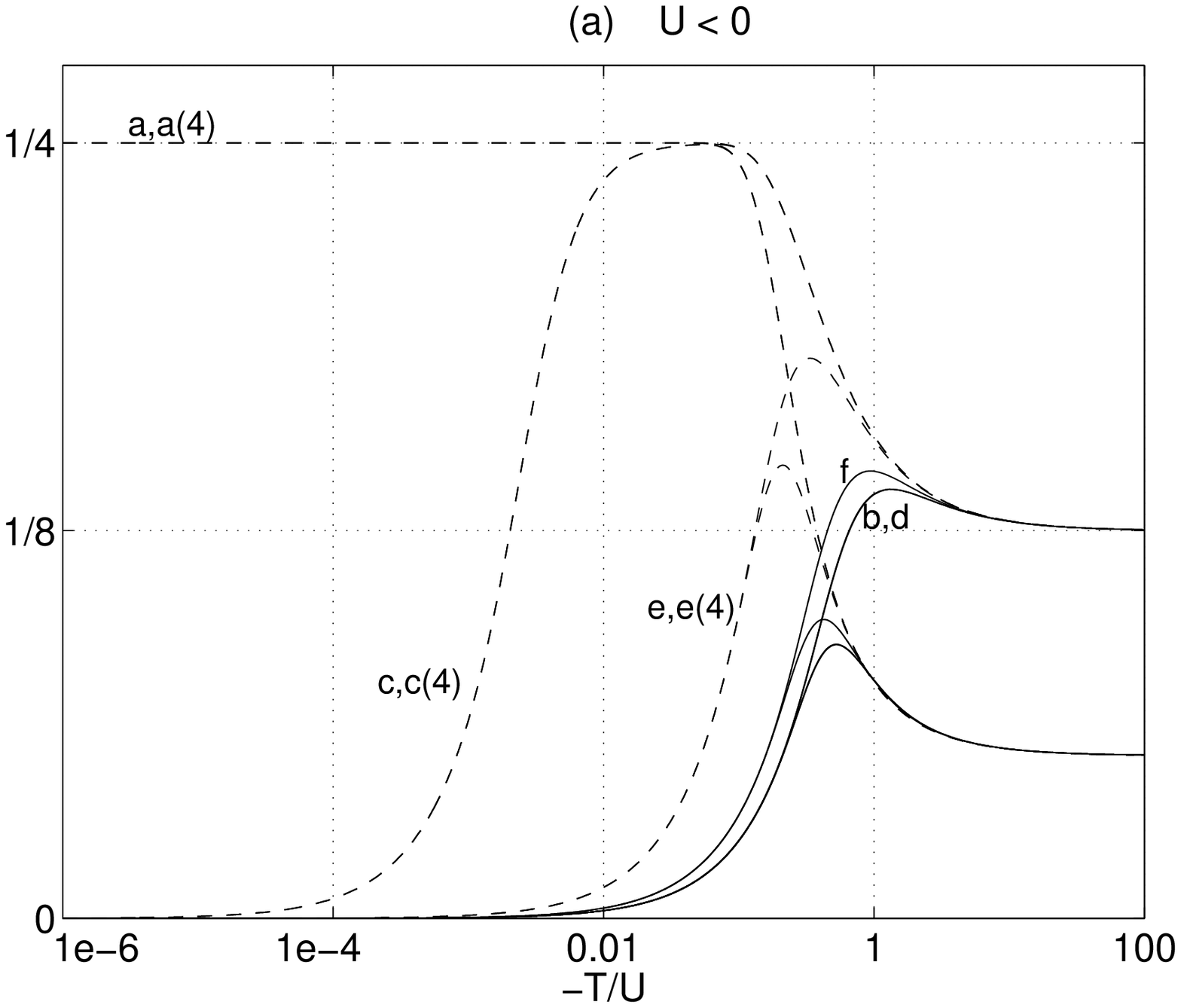,width=\columnwidth}
\end{center}
\end{minipage}
\hfill
\begin{minipage}[t]{.45\textwidth}
\begin{center}
    ~\psfig{file=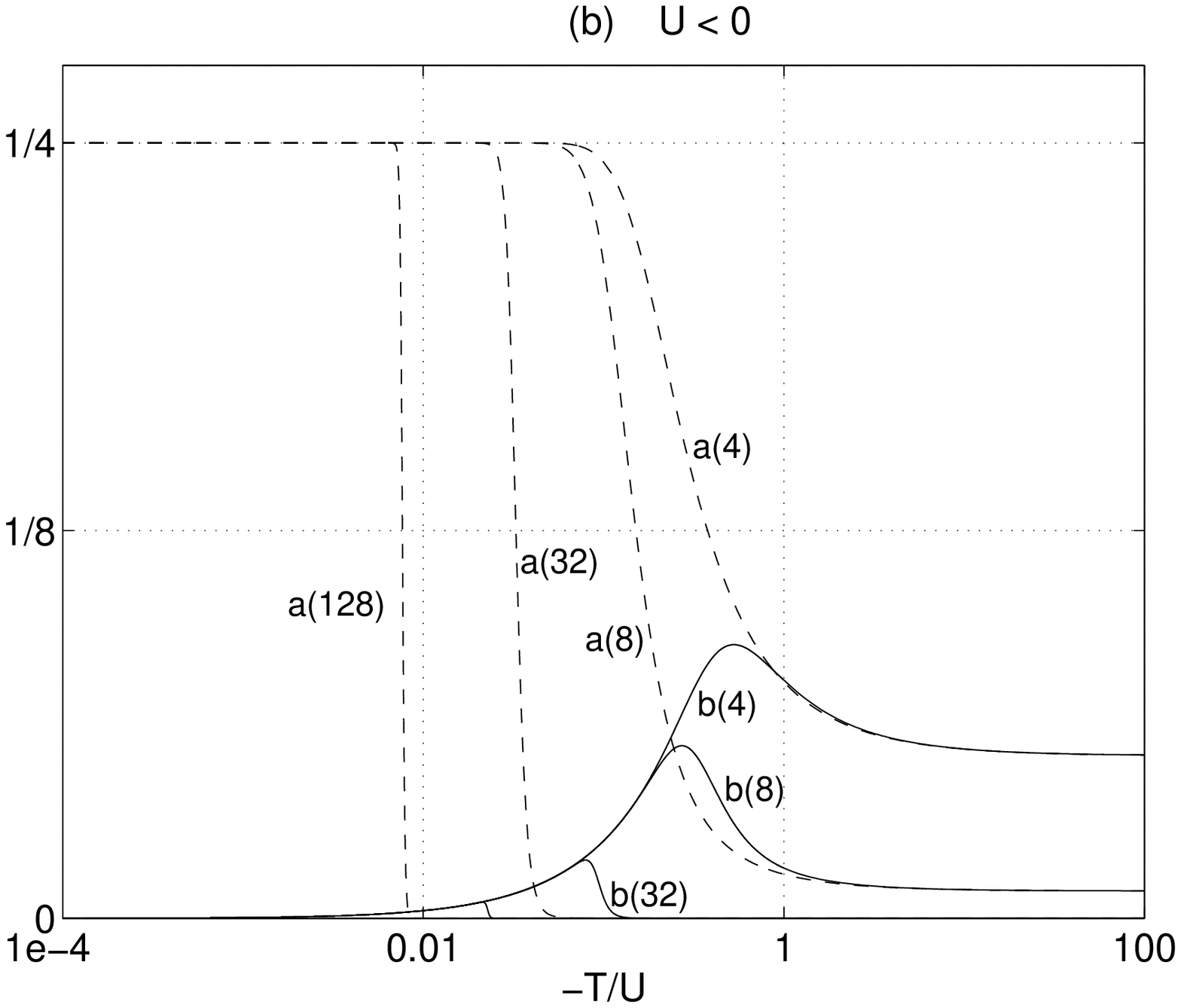,width=\columnwidth}
  \end{center}
\end{minipage}
\hfill
\caption{\label{Fig10}Field dependence of $\xi_{\perp}$, for $U<0$ in
  (a) moderate and (b) strong fields.}
\end{figure}

The charge susceptibility in zero field and for weak and strong
magnetic fields is illustrated in \ofigs\ref{Fig8} and \ref{Fig9} for
the parallel component and in \ofig\ref{Fig10} for the perpendicular
part.  The low-temperature charge Curie law $T\charsusc=1/4$ is here
found to be obeyed only in the symmetric situation ($y=-1/2$,
$\xi_\perp=\xi_\parallel$), irrespective of the magnetic field.
Furthermore, an infinitesimal asymmetry is sufficient to depress the
low-temperature limit as shown in \ofig\ref{Fig8}a, where the curve
"b"\, corresponds to $y\approx-0.499992$.  Again, in high fields a peak
of height $1/16$ is formed in the parallel charge response
$T\charsuscpar$ for $y>0$, while the perpendicular component displays
a threshold at the same point. However, here for $y=1/2$ this
behaviour corresponds to the level configuration shown in
\ofig\ref{Fig5}c, where $B=-3U$.

\begin{figure}
\hfill
\begin{minipage}[t]{.45\textwidth}
  \begin{center}
    \hspace{0.1\columnwidth}\psfig{file=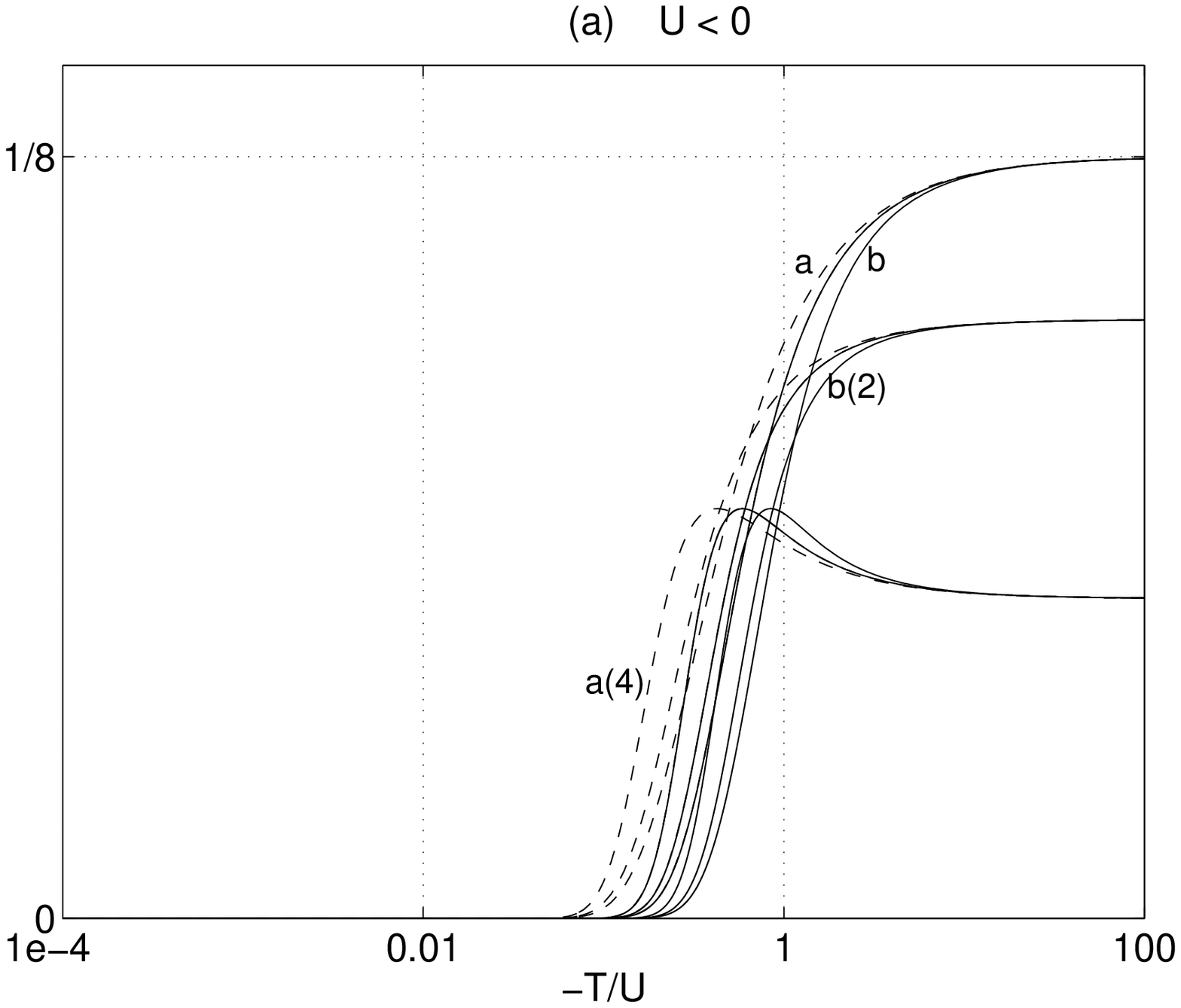,width=\columnwidth}
\end{center}
\end{minipage}
\hfill
\begin{minipage}[t]{.45\textwidth}
\begin{center}
    ~\psfig{file=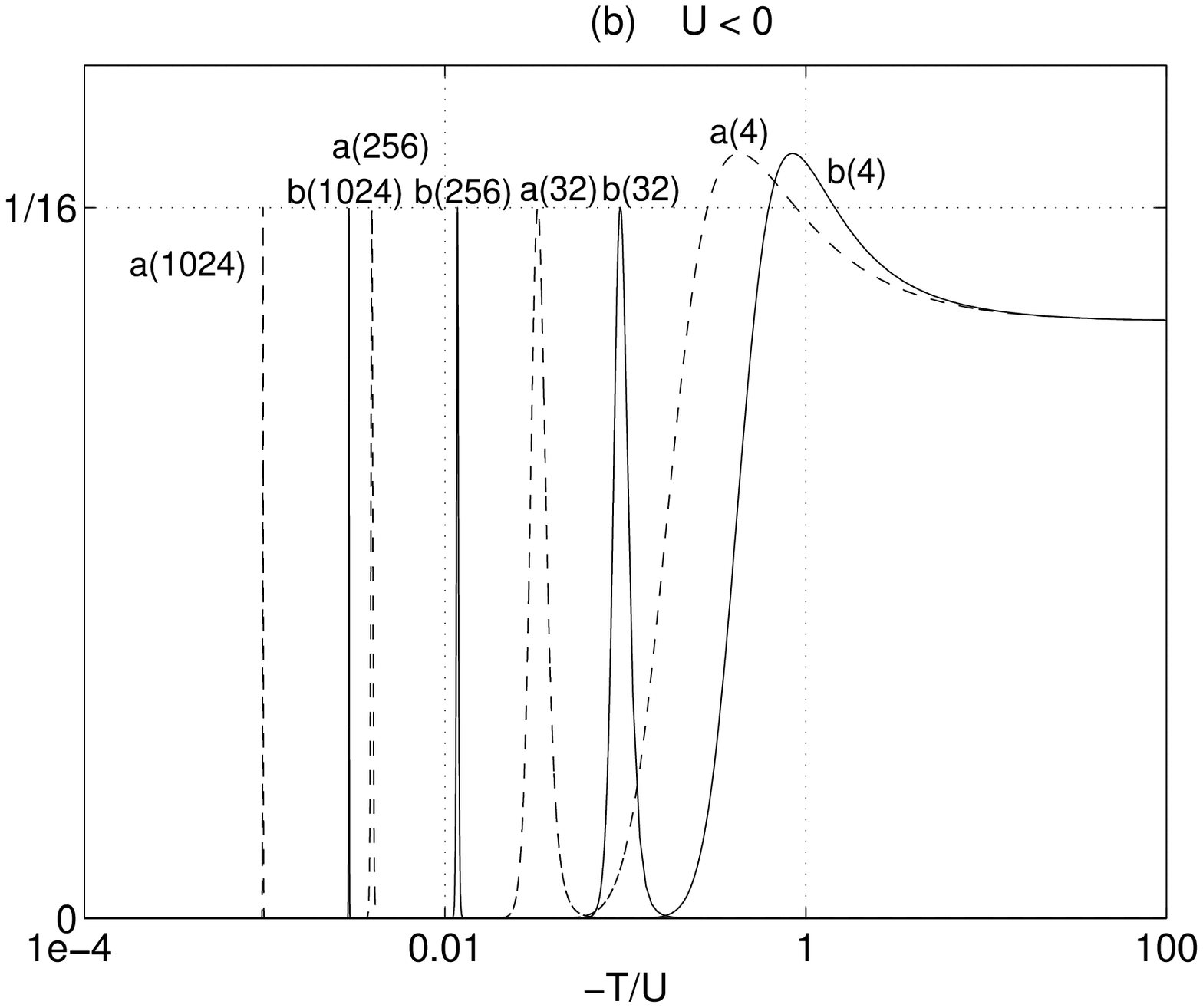,width=\columnwidth}
  \end{center}
\end{minipage}
\hfill
\caption{\label{Fig11}(a) For $U<0$, the longitudinal spin 
  susceptibility $\chi_{\parallel}$ is strongly suppressed for
  $T<U/10$.  (b) A sharp peak is formed for extreme fields for
  both positive and negative values of $y$.  Curves "a"\, and "b"\,
  are for $|y|=1/2$, corresponding to level configurations in
  \ofigs\ref{Fig5}b and \ref{Fig5}c, while the other curves in (a)
  represent $|y|=1/2^{10}$.}
\end{figure}

\begin{figure}
\hfill
\begin{minipage}[t]{.45\textwidth}
  \begin{center}
    \hspace{0.1\columnwidth}\psfig{file=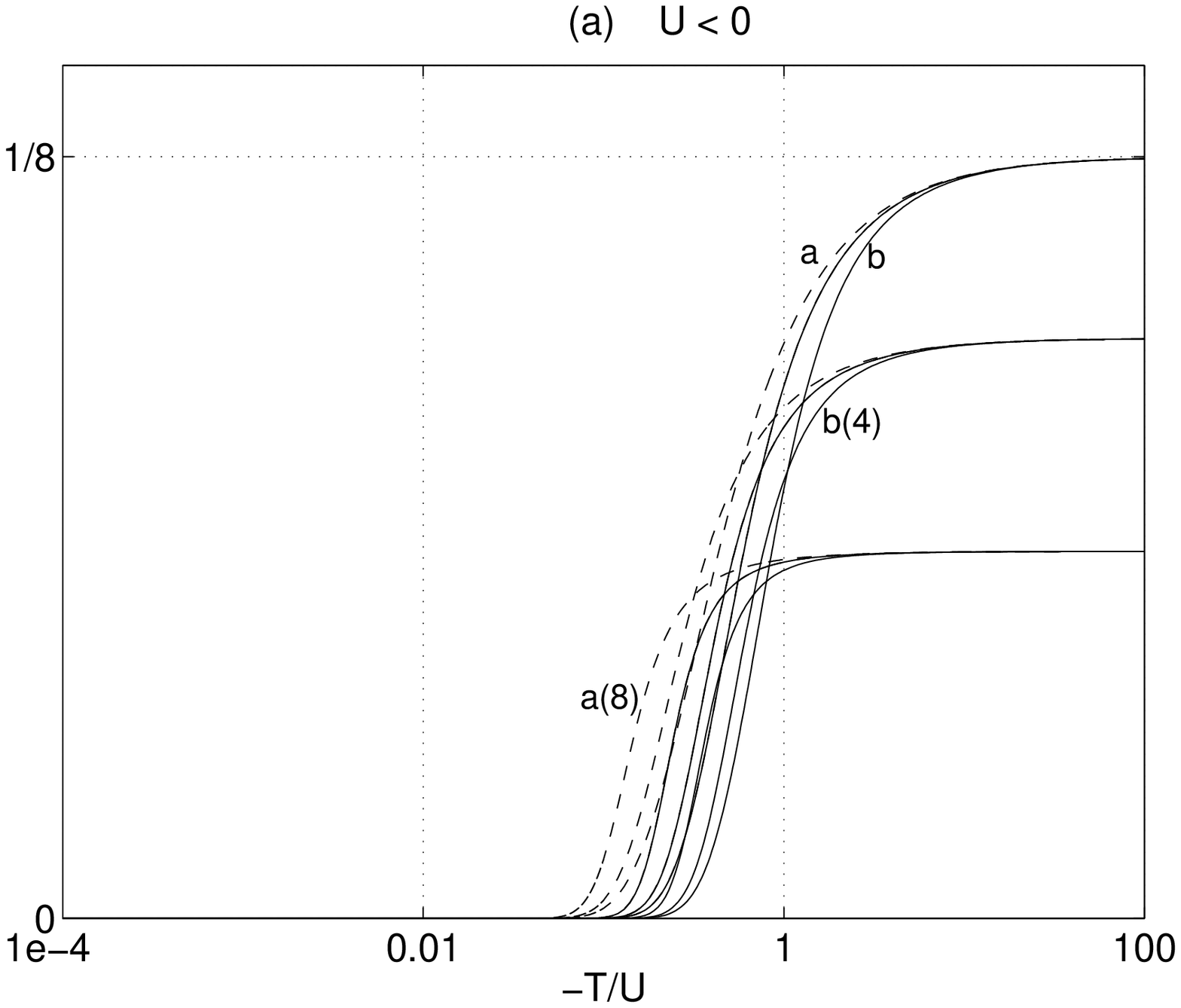,width=\columnwidth}
\end{center}
\end{minipage}
\hfill
\begin{minipage}[t]{.45\textwidth}
\begin{center}
    ~\psfig{file=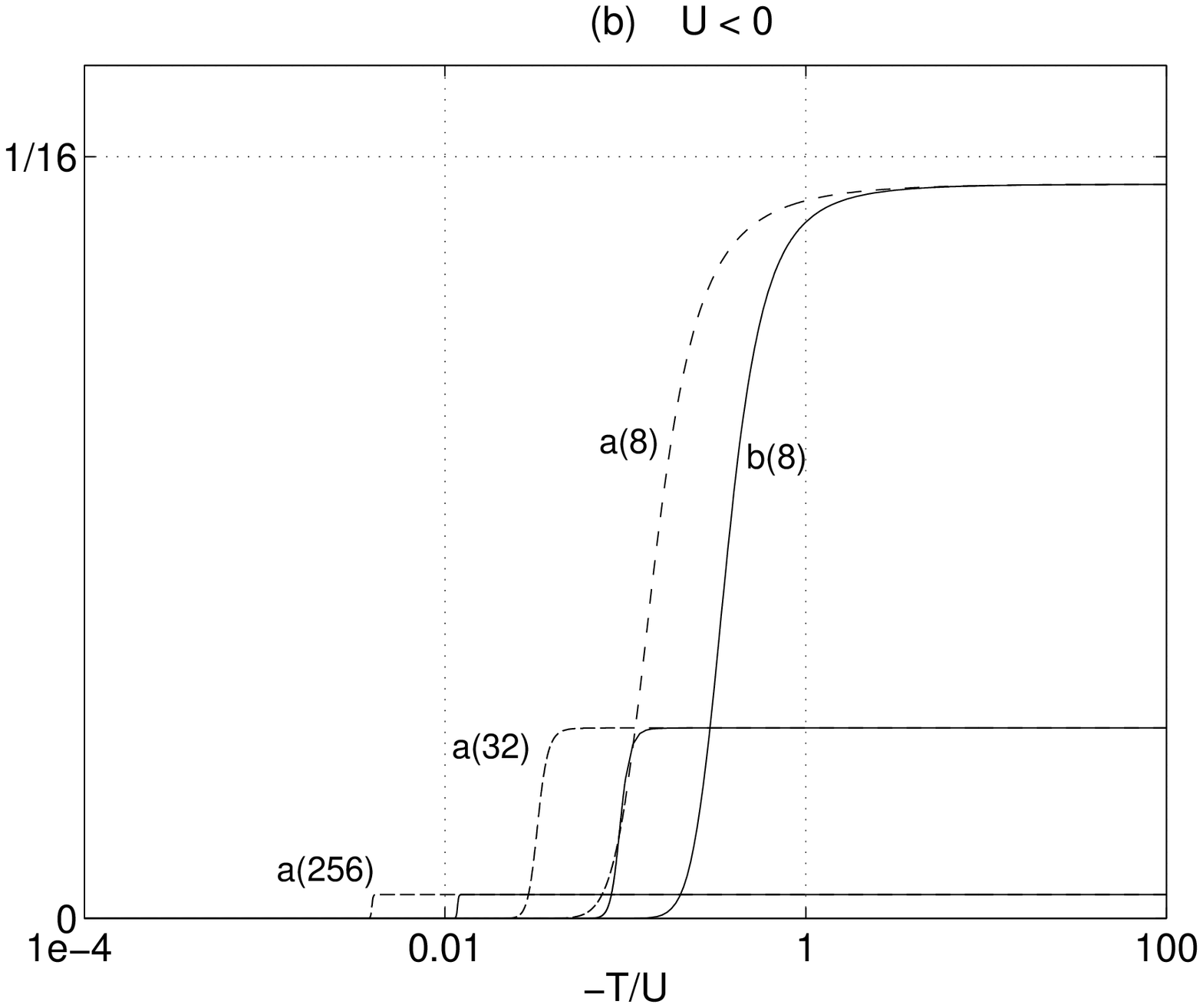,width=\columnwidth}
  \end{center}
\end{minipage}
\hfill
\caption{\label{Fig12}(a) For $U<0$, the transversal spin 
  susceptibility $\chi_{\perp}$ is strongly suppressed at low
  temperatures. (b) Extreme field values produce a threshold
  for both $y>0$ and $y<0$ corresponding to the peaks in
  \ofig\ref{Fig11}. The curves have been drawn for the same asymmetries as in
  \ofig\ref{Fig11}.}
\end{figure}

For $U<0$ the spin susceptibilities behave in a similar way as the
charge response functions for $U>0$, freezing out for low enough
temperatures, as shown in \ofigs\ref{Fig11} and \ref{Fig12}.  However,
the $y=0$ situation is not a special case here. Again we observe that
the parallel spin susceptibility is more sensitive to the external
field than the perpendicular component. Furthermore, we find peaks in
$T\spinsuscpar$ and thresholds in $T\spinsuscperp$ for high fields,
but now both at $y>0$ and $y<0$.  The $y=1/2$ peaks and thresholds in
\ofigs\ref{Fig11}b and \ref{Fig12}b correspond to the level
configuration of \ofig\ref{Fig5}c ($B=-3U$), while those for $y=-1/2$
correspond to the situation in \ofig\ref{Fig5}b, where $B=-U$.

\subsection{Contrasting Spin and Charge}
\begin{figure}
  \begin{center}
    \hspace{0.1\columnwidth}\psfig{file=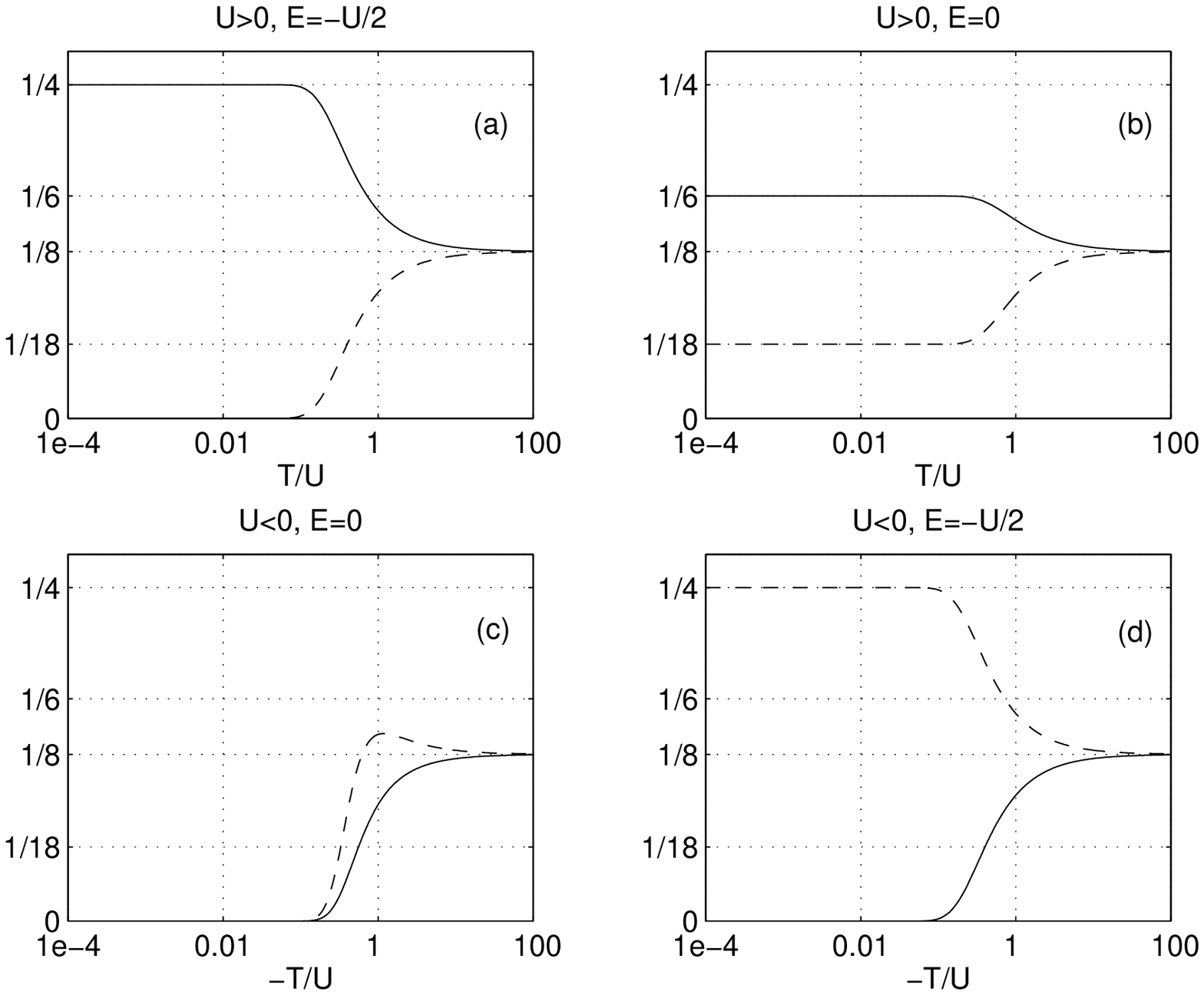,width=1.0\columnwidth}
  \end{center}
\caption{\label{Fig13}Contrasting spin ($\chi_\parallel$) 
  (solid line) and charge ($\xi_\parallel$) (dashed line) response
  functions for $U>0$ (a and b) and for $U<0$ (c and d). Note, in
  particular, the spin-charge duality: exchange in the roles of $\chi$
  and $\xi$ for $U\leftrightarrow -U$ (between a and d).  For further
  discussion, see the main body of text.}
\end{figure}

As pointed out above, the reversal of $U$ exchanges the mutual roles
of the spin and charge degrees of freedom. We find, in particular,
that at zero field for the symmetric situation ($y=-1/2$):
$\charsusc(U<0)=\spinsusc(U>0)$ and $\charsusc(U>0)=\spinsusc(U<0)$.
This is illustrated in \ofig\ref{Fig13} for the parallel
susceptibilities.  Furthermore, we find that the spin and charge
response functions are symmetric with respect to the level asymmetry
$a=2E+U$, such that $\chi(a)=\chi(-a)$ and
$\xi(a)=\xi(-a)$ are obeyed.

\section{DISCUSSION}
\label{sec:discussion}
The Anderson impurity model has proven capable of describing a
remarkable variety of different physical systems in the field of
strongly correlated electrons.  Presently, an important trend in
condensed-matter physics and nanoelectronics is one where artificial
man-made objects are studied, rather than real atoms or molecules.
The progress in lithography techniques has made it possible, e.g., to
fabricate quantum dots with properties similar to those of real atoms.
Our study has relevance for such systems where isolated localized
energy levels occur. It would be interesting to investigate whether
indications for negative-$U$ behaviour previously found, e.g., in the
context of high-temperature superconductivity and heavy-fermion
systems could also be experimentally realized in quantum dots 
\cite{schlottmann} or carbon nanotubes with magnetic impurities.

We have studied the spin and charge susceptibilities of an artificial
Anderson atom for arbitrary values of the model parameters.  General
results for all the four relevant response functions have here been
presented to the best of our knowledge for the first time and an
extensive survey of the properties of these susceptibilities was
carried out.  It was pointed out, in particular, how the level
asymmetry behaves for the charge degrees of freedom as the magnetic
field for the spin, and that the reversal of $U$ changes the mutual
roles of spin and charge.  In particular, for low temperatures and
with positive $U$, the Curie law for a free spin-$1/2$ is followed by
the spin response, whereas the charge excitations become suppressed.
On the other hand, for negative $U$ at low temperatures, the charge
susceptibility follows the Curie law in the symmetric case, whereas
the spin responses vanish.  At high temperatures and for increasing
asymmetry, the transfer of oscillator strength from the spin degrees
of freedom to the charge modes becomes increasingly important. In the
$T\rightarrow \infty$ limit, thermal fluctuations average over any
details in the energy-level structure.  Consequently, the spin and
charge modes
 have equal oscillator strengths and they become 
equally relevant.

It is noted that for increasing magnetic-field strength, the longitudinal
susceptibilities are more strongly suppressed than the perpendicular
components.  Furthermore, at extremely high fields, the longitudinal
responses show peaks of invariant height while the perpendicular
responses have a threshold. These features are associated with
particular level crossings, as discussed in the text.

Our thorough discussion of the spin and charge (pseudospin)
susceptibilities in the atomic limit of the Anderson model has
revealed several new and interesting details in these response
functions.  These new features are expected to be particularly
relevant in connection with the behaviour of quantum dots since in
this case it is possible to create very high effective magnetic fields
far beyond those encountered in atomic physics. Therefore, quantum
dots serve as interesting laboratory models for Anderson model physics
in limits which have not been explored before.  Our results serve
to emphasize and increase understanding of the
relationships governing the spin-charge duality in the atomic limit of
the Anderson model. The atomic model can be taken as the starting
point of perturbation expansions in the Schrieffer-Wolff limit where
the Anderson model can be related to the Kondo model for magnetic
impurities in metals \cite{schrieffer}. Unified definitions of spin
and charge susceptibilities have been given and they will be utilized
in future works.

\appendix
\section{RESPONSE FUNCTIONS}
\label{sec:response}
In order to describe the response of an operator $A$ due to a
perturbation coupled to operator $B$, it is natural to investigate the
generalized admittance function \cite{rickayzen}, defined in Zubarev's
\cite{zubarev} notation as
\begin{equation}
  \green{A}{B}^{(\pm)}_z=\mp\, i\!\int_{-\infty}^\infty \!\!\!\!dt\, e^{izt}\, \theta(\pm t)\, 
  \langle [A(t),B(0)]_{(\pm)} \rangle,
  \label{eq:greendef}
\end{equation} 
which is a complex function of the frequency variable $z$ and has been
analytically continued for complex arguments. The upper and lower
signs designate ${\rm Im}(z) > 0$ and ${\rm Im}(z) < 0$, respectively,
and correspond to the retarded (analytic in the upper half of the
complex plane) and advanced (analytic in the lower halfplane)
functions. The superscript $(+)$ refers to an anticommutator
($[A,B]_{(+)}\equiv\{A,B\}$, correlation) function, while a commutator
($[A,B]_{(-)}\equiv[A,B]$, response) function is meant with the
superscript $(-)$. Do not confuse $(\pm)$ (signs in parentheses)
marking the commutator $(-)$ and anticommutator $(+)$ functions with
$\pm$ or $\mp$ (without parentheses) denoting the retarded and
advanced functions in \protect\eqref{eq:greendef}.

The time evolution of operators is ruled in the Heisenberg
representation by the operator equation
\begin{equation}
  A(t)=e^{i\ham t}\, A(0)\, e^{-i\ham t},
  \label{eq:operheis}
\end{equation}
where $\ham$ is the Hamiltonian.  Partial integration of 
(\ref{eq:greendef}) yields with the help of (\ref{eq:operheis})
the equations of motion
\begin{eqnarray}
  z\green{A}{B}^{(\pm)}_z &=&\langle [A,B]_{(\pm)}\rangle + 
  \green{[A,\ham]}{B}^{(\pm)}_z
  \label{eq:greenmotion1}\\
  &=&\langle [A,B]_{(\pm)}\rangle - \green{A}{[B,\ham]}^{(\pm)}_z,
  \label{eq:greenmotion2} 
\end{eqnarray}
where the ensemble averages $\langle\rangle$ can be
related selfconsistently to the respective admittance functions (see
below).  For most of the nontrivial Hamiltonians of interest, the
equations of motion for the relevant double-time Greens functions
constitute an infinite hierarchy, whose termination ({\it e.g.}, in
the Hartree-Fock approximation) has been a frequently employed
approximation technique in the theory of magnetism. It is therefore of
interest to consider models where the exact double-time functions can
in fact be evaluated in closed form.

For large frequencies, the double-time functions decrease at least as
$\green{A}{B}\sim z^{-1}$, as can be readily seen from their equations
of motion (\ref{eq:greenmotion1}) and (\ref{eq:greenmotion2}).
Moreover, these functions are analytic off the real frequency
$(\omega)$ axis, across which there exists a cut discontinuity
\begin{equation}
  \green{A}{B}_{\omega\pm i0}=\green{A}{B}_{\omega}^{'}\pm i\green{A}{B}_{\omega}^{''},
  \label{eq:greendisc}
\end{equation}
where the prime superscript denotes the real part, while the double
prime stands for the imaginary part of the function.  Thus the
response functions may be represented using the Hilbert transformation
\begin{equation}
  \green{A}{B}_z=\int\frac{d\omega}{\pi}\frac{\green{A}{B}^{''}_{\omega}}{\omega -z},
  \label{eq:hilbert}
\end{equation}
which is also called spectral representation.

For the anticommutator $(+)$ metric, the time-correlation functions
(expectation values) may be obtained using
\begin{eqnarray}
  \langle \delta A(t)\,\delta B(0)\rangle & = & \int \frac{\D
    d\omega}{\D\pi}\,e^{-i\omega t}\,\left[1-f(\omega)\right]\,
  G_{AB}^{''}(\omega)
  \label{eq:corr1} \\
  \langle \delta B(0)\,\delta A(t)\rangle & = & \int \frac{\D
  d\omega}{\D\pi}\,e^{-i\omega t}\,f(\omega)\,G_{AB}^{''}(\omega),
  \label{eq:corr2}
\end{eqnarray}
where $\delta A = A - \langle A \rangle$, $\delta B = B -
\langle B \rangle$, $G_{AB}(z)=-\green{A}{B}_z^+$ and $f(\omega)$ is
the Fermi distribution function (temperature units are chosen such
that the Boltzmann constant is unity). For the commutator
(response, $(-)$) functions we have 
\begin{eqnarray}
  \langle \delta A(t)\,\delta B(0)\rangle&=&\int\frac{\D
  d\omega}{\D\pi}\,e^{-i\omega t}\,\left[1+n(\omega)\right]\,\chi_{AB}^{''}(\omega)
  \label{eq:corrchi1}\\
  \langle \delta B(0)\,\delta A(t)\rangle&=&\int\frac{\D
  d\omega}{\D\pi}\,e^{-i\omega t}\,n(\omega)\,\chi_{AB}^{''}(\omega),
  \label{eq:corrchi2}
\end{eqnarray}
where the $A-B$ susceptibility is defined as
$\chi_{AB}(z)=-\green{A}{B}_z^-$ and $n(\omega)$ is the Bose function.
Furthermore, the fluctuation-dissipation theorem
\begin{equation}
  G_{AB}^{''}(\omega)=\coth\left( \frac{\omega}{2T} 
  \right)\chi_{AB}^{''}(\omega)
  \label{eq:flucdis}
\end{equation}  
gives a relation between the commutator and anticommutator functions.

%
%

\section*{ACKNOWLEDGMENTS}
Part of this work was carried out at the Institute f\"ur Theorie
der Kondensierten Materie at the Universit\"at Karlsruhe,
Germany. MMS is grateful to P.~W\"olfle for cordial hospitality,
G.~Sch\"on for discussions and the HERAEUS Foundation for financial
support. MPVS acknowledges the support by the Finnish Cultural 
Foundation. This work has also been supported  by the Academy of
Finland through the project "Theoretical Materials Physics".

\end{document}